\documentclass{aa}  

\usepackage{graphicx}
\usepackage{txfonts}
\usepackage{textcomp}
\usepackage[colorlinks=true, linkcolor=blue, citecolor=blue,
urlcolor=blue, breaklinks=true]{hyperref}
\newcommand{\tnotecaption}[1]{\\ \vspace*{0.5\abovecaptionskip}{\normalfont\small\bfseries Notes:~}#1} 

\usepackage{color}
\usepackage[normalem]{ulem}

\begin{document} 
    
   \title{Orbital obliquity sampling in the Kepler-20 system using the 3D animation software Blender} 
   
   \author{H.\,M. M\"uller, P. Ioannidis, \and J.\,H.\,M.\,M.\, Schmitt}
\authorrunning{M\"uller et al.}%  

   \institute{Hamburger Sternwarte, Universit\"at Hamburg, Gojenbergsweg 112, 21029
Hamburg, Germany\\ \email{hmueller@hs.uni-hamburg.de}}

   \date{Received 24.06.2021; accepted 03.08.2021}
% accepted 03.08.2021

% \abstract{}{}{}{}{} 
% 5 {} token are mandatory
 
  \abstract
  % context heading (optional)
  % {} leave it empty if necessary  
   {The mutual orbital alignment in multiple planetary systems is an important parameter for understanding their formation.
   There are a number of elaborate techniques to determine the alignment parameters using photometric or spectroscopic data.
   Planet--planet occultations (PPOs), which can occur in multiple transiting systems, are one intuitive example.
   While the presence of PPOs constrains the orbital alignment, the absence at first glance does not.}
  % aims heading (mandatory)
   {Planetary systems, for which the measurement of orbital obliquities with conventional techniques remains elusive, call 
   for new methods whereby at least some information on the alignments can be obtained.
   Here we develop a method that uses photometric data to gain this kind of information from multi-transit events.}
  % methods heading (mandatory)
   {In our approach we synthesize multi-transit light curves of the exoplanets in question via the construction
   of a grid of projected orbital tilt angles $\alpha$, while keeping all transit parameters constant.
   These model light curves contain PPOs for some values of $\alpha$.   To compute the model light curves,
   we use the 3D animation software Blender for our transit simulations, which allows the use of arbitrary surface brightness distributions 
   of the star, such as limb darkening from model atmospheres. The resulting model light curves are then compared to actual measurements.}
  % results heading (mandatory)
   {We present a detailed study of the multi-transiting planetary system \object{Kepler-20}, including parameter fits of the transiting planets and an analysis of the stellar activity.
   We apply our method to Kepler-20\,b and c, where we are able to exclude some orbital geometries, and find a tendency of these planets to eclipse in front of different stellar hemispheres in a prograde direction.}
  % conclusions 
   {Despite the low statistical significance of our results in the case of Kepler-20, we argue that our method is valuable for systems where PPO signals larger than the noise can occur.
   According to our analysis, noise $\le 2\cdot10^{-4}$ for planets like Kepler-20\,b, or a planet radius $\ge 3\,R_\mathrm{Earth}$ for the smaller component and Kepler-20-like photometry, would 
   be sufficient to achieve significant results.}

%Stars: individual: Kepler-20
   
   \maketitle
%
%-------------------------------------------------------------------

\section{Introduction}
\label{sec:intro}

A significant fraction of the exoplanets discovered so far are located in planetary systems, which means the host star is orbited by two or more planets.
At least one-third of the confirmed transiting exoplanets are hosted in 791 multiple systems\footnote{\url{exoplanet.eu}, as of August 2021}.
The discovery and analysis of such systems are of special interest for exploration of the diversity of exoplanetary systems and to test planet formation theory.
For these studies, all the parameters of the systems' geometry, as in the semi-major axes and the orientation of the orbital planes with respect to the stellar rotation axis, $\psi$, have to be known.
However, the latter, often denoted the spin--orbit misalignment, is difficult to measure.
Thus, in most cases, one measures only the projected spin--orbit angle, $\lambda$, which is the angle between the sky projections of the angular momentum vectors, and allows the determination of a lower limit on $\psi$ \citep{winn2009}.
Together with a growing number of measured obliquities, this will also allow us to set our Solar System in the larger context of multiple-planet systems. 

Several approaches exist for the determination of the orbital obliquities of exoplanets.
The most commonly used is based on the Rossiter-McLaughlin effect \citep{Rossiter1924,McLaughlin1924}.
Developed for the study of eclipsing binaries, this method measures an apparent Doppler shift induced by light blocked from specific parts of the stellar surface, which holds velocity information because of the stellar rotation.   
In this way, this method is sensitive to the projected spin--orbit angle $\lambda$.
With the increasing performance of the instrumentation used, this method has been  adopted for the study of transiting exoplanets, as described by \citet{Ohta2005} and \citet{Hirano2011}.
Driven by the efforts of numerous authors, this method has become the most successful technique for measuring orbital obliquities in exoplanet systems, even at chromospheric wavelengths \citep{Czesla2012} or in the case of a young and active M-dwarf as planet host \citep{Martioli2020}.

However, this method has its limitations when it comes to rapidly rotating massive stars.
Because of their fast rotation, these stars show significant oblateness and gravity darkening, both observable in transit photometry.
As introduced by \citet{Barnes2009}, these effects offer the opportunity to measure the orbital alignment by transit modeling.
\citet{Barnes2011} present the first measurement of this kind using data of a gravity-darkened host star.
In this case, the stellar disk brightness distribution scanned by a transiting planet holds information about the orbital alignment, which is also provided by activity indicators, such as spots.
If individual spots have been identified and found to occur repeatedly in transits, one can assume an alignment of the transit path with the spot latitudes of the host star, as shown in the work of \citet{Desert2011}.

The Doppler tomography presented by \citet{Collier2010} is a combination of modeling the missing light during the transit and the projected radial velocities sampled by the planet, which has also been applied in an increasing number of cases to determine spin--orbit alignment \citep[e.g.,][]{Martinez2020}.
In contrast, to analyze only the time-frame during which the planetary transit occurs, 
it is possible to take long-term photometry into account.
This is, for example, the case for the statistical analysis of the rotational amplitude distribution given by \citet{Mazeh2015}.
In their work, these latter authors find a relation between spot modulation and the inclination of the stellar rotation axis.
For the case of transiting planets, \citet{Mazeh2015} highlight indications for a spin--orbit alignment of cool stars and a corresponding misalignment for hot stars.
Instead of using rotational amplitudes to determine the stellar inclination and gain alignment information, it is also possible to use the stellar rotation period combined with measurements of the rotational velocity \citep[see, e.g.,][]{Morton2014,Masuda2020}.

Another promising method using long-term photometry is based on asteroseismology, such as that presented by \citet{Campante2016}.
By fitting the oscillation modes, it is also possible to measure the stellar inclination, which together with the orbital inclination and $\lambda$ allows the spin--orbit angle $\psi$ to be statistically constrained.
In addition, all of these methods are also valid for multi-planet systems, and so these systems can hold additional information about the mutual alignments for example.
Because of the gravitational interaction of these planets, it is possible to observe transit timing variations (TTVs) or even transit duration variations (TDVs); e.g.,  \citet{Nesvorny2013}.
Dynamical orbit fits to these variations allow  the set of orbital elements of the involved planets to be to determined.
These, together with inclination measurements of the stellar rotation axis, can be used to constrain the spin--orbit alignment \citep[e.g.,][]{Huber2013}.
Additionally, such multiple systems are capable of producing multi-transit events, that is, transits where two (or more) planets eclipse their host star at the same time.
It might be possible that two planets, which perform a multi-transit, can mutually overlap during the event, caused by different orbital velocities.
Such an event is called a planet--planet occultation (PPO) and is visible as a bump in the multi-transit light curve \citep[e.g.,][]{Pal2012}.
\citet{Hirano2012} were the first to present the discovery of such a rare event in the \object{Kepler-89} system.
Such a PPO is indeed only possible in the case of matching orbital alignments.
This implies that in the case of the detection of PPOs, the orbital geometry ---including time information like orbital phases--- can be constrained to a considerable degree; see for example \citet{Masuda_2013} for Kepler-89 d and e, and \citet{Luger2017} for TRAPPIST-1 and other systems.

While these approaches were used to measure $\lambda$ in only 142 cases\footnote{Crossmatch of \url{exoplanet.eu} and \url{exoplanets.org}, as off August 2021} ---providing an accuracy ranging from $\pm0.3\degr$ \citep{Triaud2009} to $\pm79.5\degr$ \citep{Eastman2016}---,the orbital alignment of the vast majority of exoplanets is still unknown.
This relatively low number is probably caused by the fact that these measurements require high-quality photometric and time-resolved spectral observations, which are quite hard to achieve for visually faint exoplanet host stars. 
Additional methods to gain insight into the orbital geometry are therefore of special interest.

In the present paper we introduce a new method to determine orbital alignment information in transiting multi-planet systems that only relies on photometric data.
For that purpose, we first explain our approach in Sect.~\ref{sec:methods}, which was previously presented in a preliminary form as a poster presentation \citep{meinPoster2019}.
We also comment on different geometric cases as starting conditions and present a Bayesian approach to distinguish between them.
We introduce the software we use and point out its advantages.
We then apply our method to real data, present and discuss the results, and we end with a summary and our conclusions.

\section{Methods}
\label{sec:methods}

\subsection{Orbital obliquity sampling}
\label{sec:oom}
\begin{figure}
\centering
\includegraphics[width=0.5\textwidth]{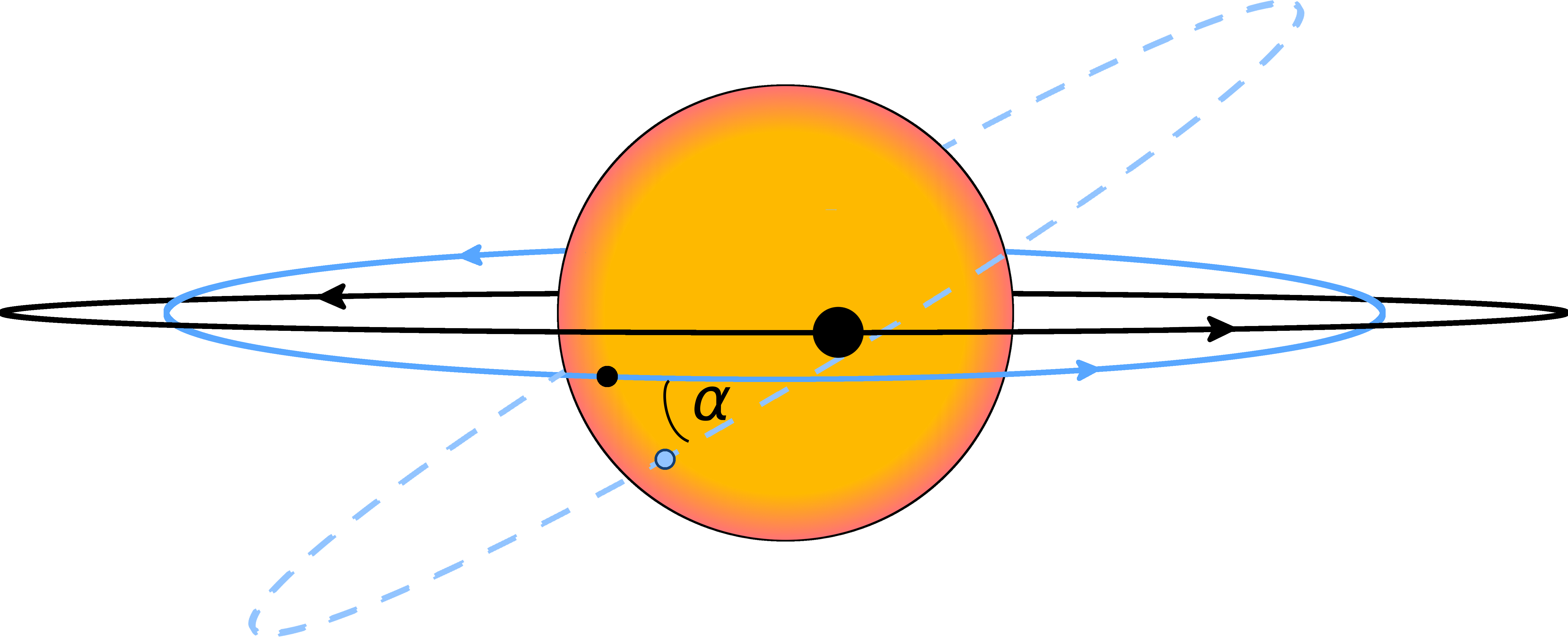}
 \caption{Illustration of our grid simulation.
The orbit of planet b is rotated by the angle $\alpha$ (\citet{meinPoster2019}, Fig.~5).
In the new orbit geometry (\textit{dashed line}) a PPO can occur. Orbital movements are indicated by arrows.}
  \label{fig:method} 
\end{figure}

Multi-transiting systems offer the opportunity to observe multiple transits or even PPOs, TTVs, or TDVs, which can be used to determine orbital alignment data.
While PPOs tend to be exceedingly rare because of the requirement of matching orbital geometries and phases, TTVs and TDVs are not always sufficiently pronounced to be significant.
We argue that something can be learnt about the orbital geometries even if the mentioned properties are not present or are not significant in the data.
For that purpose, we pursue the idea of comparing simulated multiple transits to real data, requiring
that these simulations are carried out for a grid of different orbital alignments.
In Fig.~\ref{fig:method} we illustrate our approach to achieve this grid.
While two planets are on their orbital tracks performing a multiple transit, we tilt one of the orbits by the angle $\alpha$.
After one of these events is complete we increase $\alpha$ by $1\degr$ and repeat the simulation.
This process is performed for all possible angles.
In this way we are able to achieve simulated PPOs for individual angles, which leave their imprint on the model light curve.
The comparison between model and real data is then expressed as a $\chi^2$-value as a function of the simulated tilt angle $\alpha$.
This $\chi^2(\alpha)$ curve can then be used as a statistical estimator to decipher the tilt angles that are favorable for the system and those that are not.
A comparable statistical approach is presented by \citet{Sanchis2011}, who rely on star spots and present a different analysis.

\subsubsection{Multiple transits and planet--planet occultations}
Simulated multiple transit light curves can easily be generated by calculating the product of two or more individual artificial transit light curves based on different orbital and planetary parameters.
This can be achieved, for example, by using the \texttt{occultquad}\footnote{\url{http://www.astro.washington.edu/users/agol}} routine \citep{MandelAgol2002} and ending up with light curves $\mathcal{F}_i(t)$.
Combining the individual light curves according to the relation 
\begin{equation}
    \mathcal{F}(t) = \prod_i^N \mathcal{F}_i(t) \,,\label{eq:multi_mod}
\end{equation}
where $N$ is the total number of transiting planets in the system, leads to multiple transits when the individual transits coincide in time. 

While this is the most straightforward approach to achieving multiple transits, it is not capable of creating artificial PPOs.
For this purpose, a more complex formalism is needed, such as those introduced by \citet{Pal2012} or \citet{Masuda_2013}.
The flux excess caused by a PPO event can lead to a maximum peak value that corresponds to the overlapping area of the involved 
smaller planet, that is, the absolute value is equal to its transit depth.
The duration, that is, the second most important value characterizing a PPO, depends on the projected orbital velocities relative to each other, which indeed correlates with the angle at which the orbits intersect as mentioned by \citet{Pal2012}.
As a rule of thumb, the maximum duration can be estimated by the difference between the individual transit durations.
Clearly, this neglects the orbital inclinations and the planet radii, but can still be used as an upper limit.
A comprehensive mathematical description of how to translate peak height, central time, and duration of the event into the mutual orbital alignment is given by \citet{Masuda_2013}.

When considering our Solar System, PPOs (i.e. eclipses) could also be observed by an external observer.
For instance, if the line of sight were aligned toward one ascending node, it would  be possible to observe a PPO of the Earth and the planet in question in front of the Sun.
The tilt angle between these orbits would be the orbital inclination, $i_\mathrm{SSO}$, which is by definition measured to the ecliptic as reference plane. 
For comparison, we list some examples for the Solar System  in Table~\ref{tab:list_hb}.\begin{table}
\renewcommand{\arraystretch}{1.25}
\caption[]{Orbital inclinations of some Solar System objects (SSO).}
\label{tab:list_hb}
\centering
\begin{tabular}{c|r||c|r}
\hline\hline
Object & $i_\mathrm{SSO} / \degr$   & Object & $i_\mathrm{SSO} / \degr$ \\
\hline
Mercury &7.0050 & Jupiter & 1.3044 \\
Venus   &3.3947 & Saturn  & 2.4860 \\
Earth   &0.0000 & Uranus  & 0.7726 \\
Mars    &1.8497 & Neptune & 1.7700 \\
\hline
Pluto   &17.0967& Makemake & 28.9834\\
Eris    &44.0393& Sedna    & 11.9307 \\
\hline
\end{tabular}
\tnotecaption{Values of the planets taken from \url{ssd.jpl.nasa.gov}, and of the dwarf planets from \url{minorplanetcenter.net}.}
\end{table}
We mention that our tilt angle $\alpha$ is comparable to the longitude of the ascending node, $\Omega$.
As one defines the sky-plane as reference  for exoplanet systems, there is no absolute measure of $\Omega$ available, but rather a relative measurement, $\Delta\Omega$, in relation to a second orbit.
In this scheme, we define $\alpha = \Delta\Omega$, which is similar to
$i_\mathrm{SSO}$ in the example of observable Solar System PPOs as given in Table~\ref{tab:list_hb}.

A third important aspect is the PPO frequency of any two given planets. 
If a PPO event repeats after only several orbital periods, we can confirm this event and/or measure small variations in the time of the PPO occurrence and its duration.
This is possible in the case of strong orbital resonances, which means commensurability between the orbital periods.
Such cases allow us to observe several PPOs of the involved planets in the same geometric constellation in a relatively small time frame.
In contrast, in systems with no or only weak resonances, the PPO will be most likely only a singular event or will reappear after a very long time. 
In the context of our obliquity sampling it is of interest which case is applicable.
In the resonant example, only a severely limited range of angles $\alpha$ would lead to PPOs, always the same for every observed multiple transit. In the other case, all possible geometries have to be considered because the planets will meet at different phase angels, as these vary from multiple transit to multiple transit.

\subsubsection{Initial conditions}
\begin{figure}
\centering
\includegraphics[width=0.5\textwidth]{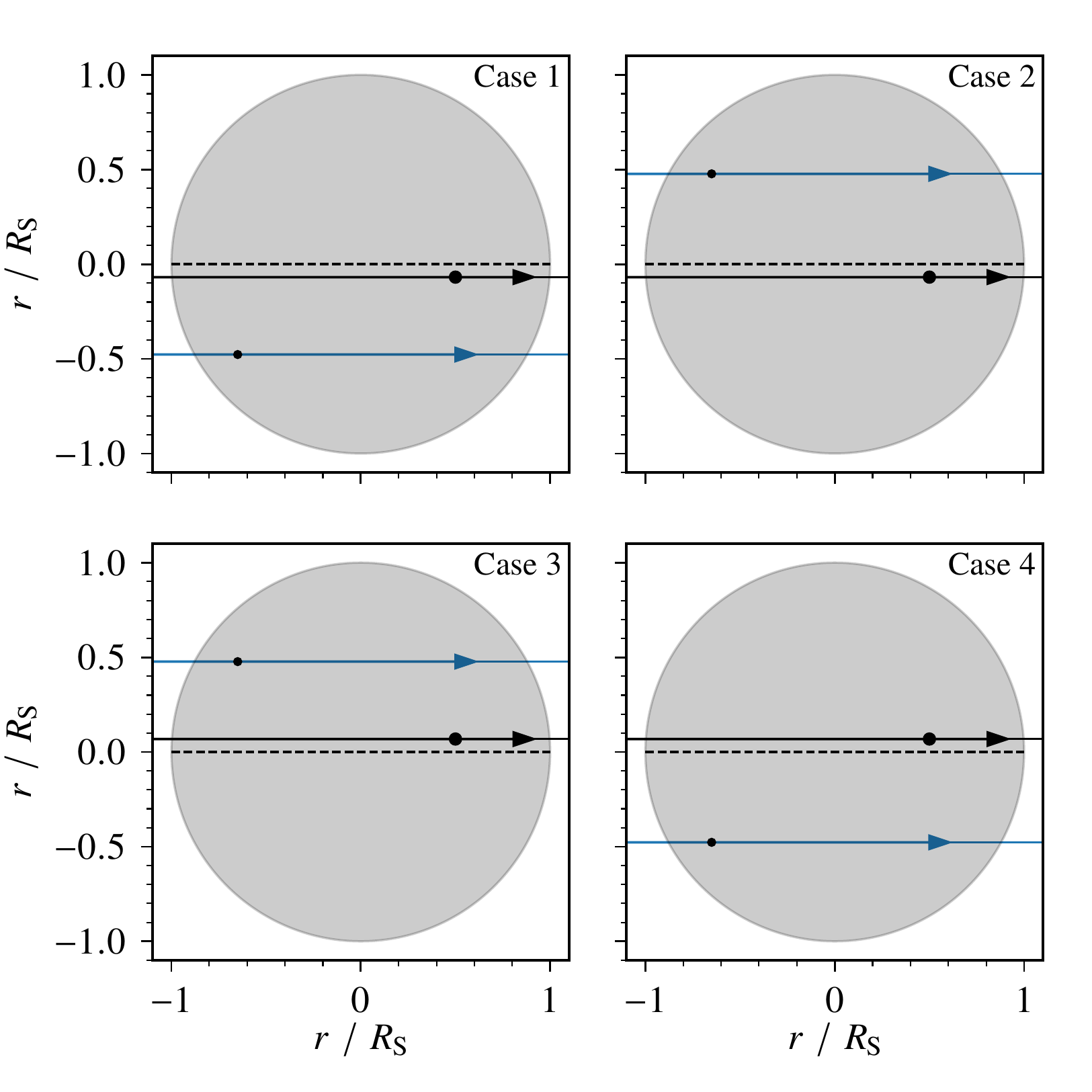}
 \caption{Illustration of four different impact parameter configurations of two planets (\textit{dots}), performing a multiple transit in front of a star (\textit{gray disk}).
 Plane parallel orbits are shown as \textit{solid} lines with the directions of motion indicated by \textit{arrows}, while the stellar equator is marked with a \textit{dashed} line.}
  \label{fig:cases} 
\end{figure}

The transit parameters required for the simulations, namely the semi-major axis $a/R_\mathrm{S}$, the orbit inclination $i$, the planet-to-star radius ratio $p$, the orbital period $P_\mathrm{Orb}$, and the time of the first transit center $T_0$, are constrained by transit fits of the objects in question.
This implies that the system impact parameters of the $j$-th planet, $b_j$, are always kept fixed in our grid simulation.
However, whether the planets both eclipse their host star on the same hemisphere (or on different ones) or whether one orbit is in the retrograde direction cannot be deduced from the transits alone.
Therefore, we have to consider different starting conditions in our grid simulation that include the orbits on different hemispheres and in different directions.
In the case of two planets (e.g., b and c) causing a multiple transit, we identify four different cases, as shown in Fig.~\ref{fig:cases}, plus the same in the retrograde direction.
The orbit shown in blue indicates the one that will be rotated counter-clockwise by the tilt angle $\alpha$ to create one complete model grid.
This is done in parallel for all multiple transits of these planets available in the data set to which the models will be compared to.
We note that retrograde orbits do not have to be considered as starting conditions.
Due to the rotational symmetry of the problem, these simulations would only have to be shifted by $-180\degr$, and are already included in the prograde cases shown in Fig.~\ref{fig:cases}.
Furthermore, we identify cases 1 and 3 and cases 2 and 4 to be the same, again due to rotational symmetry.
This means that if  $\alpha$ is increased in case 1, we have to decrease $\alpha$ in case 3 by the same amount to obtain the same geometry.
The relation transforming the tilt angles of cases 1 and 2, $\alpha_{1,2}$, into those of 3 and 4 is 
\begin{equation}
 \alpha_{3,4} = 180\degr-(\alpha_{1,2}-180\degr).\label{eq:alpha_cases}
\end{equation}
Therefore, there are two different cases in total, which have to be simulated separately, namely where both planets initially start their transits in front of the same hemisphere (e.g., case 1) and in front of different hemispheres (e.g., case 4).
The simulation of the tilt angle of the black orbit is also not necessary, because a clockwise rotation of that orbit can be expressed by a counterclockwise rotation of the blue orbit, which is the same as Eq.~\ref{eq:alpha_cases}.

Indeed, the assumed symmetries and congruent cases are only valid for a rotationally symmetric stellar disk brightness distribution, that is, with no spots or gravitational darkening, and for circular orbits where the point around which the orbit is rotated lies in the center of the star, always keeping the impact parameter constant.
We note that in principle a multiple transit can also be caused by three planets; however, this is an extremely  rare case, and at best only a singular event in the available photometry.
Thus, it is reasonable to limit our investigations to multiple transits caused by two planets.

To lower the computational effort, it is reasonable to only simulate angles under which both orbits are able to actually intersect.
These angles are affected by the impact parameters $b_j$ and by the planet sizes $p_j$.
We denote the smallest angle under which PPOs can occur as $\alpha_a$.
Until the tilt angle reaches $\alpha_b$, both orbits intersect in front of the stellar disk, while for values larger than that the geometry enters the plane parallel retrograde configuration where no PPOs are possible until $\alpha_c$ is reached.
$\alpha_d$ is therefore the largest angle for which PPOs can occur in the data. 
We obtain expressions for  $\alpha_a$ to $\alpha_d$ as follows:
\begin{equation}
 \alpha_a = \sqrt{\left(\sin^{-1}(b_\mathrm{c}+p_\mathrm{c}) - \sin^{-1}(b_\mathrm{b} - p_\mathrm{b}) \right)^2} -1\degr\,, \label{eq:a_start}
\end{equation}
\begin{equation}
 \alpha_b = 180\degr - \sqrt{\left(\sin^{-1}(b_\mathrm{c}-p_\mathrm{c}) + \sin^{-1}(b_\mathrm{b}-p_\mathrm{b})\right)^2 } +1\degr\,, \label{eq:a_end}
\end{equation}
\begin{equation}
 \alpha_c = 180\degr + (180\degr-\alpha_\mathrm{b})\,,
\end{equation}
\begin{equation}
 \alpha_d = 180\degr + (180\degr-\alpha_\mathrm{a}). \label{eq:end}
\end{equation}
To cover the boundary values of $\alpha$ as integers after rounding them, we subtract $1\degr$ in Eq.~\ref{eq:a_start}, while we add $1\degr$ in Eq.~\ref{eq:a_end}.
We note that our model grid will start at $\alpha = 0\degr$ to ensure that we cover a satisfactory number of angles where no photometric interaction takes place, so that we can determine the $\chi^2$ ground level in our simulation (see our results section for more information).

\subsubsection{Bayesian statistics: How to distinguish between pro- and retrograde orbits}
\label{sect:posterior}
We define the probability that the orbital tilt angle $\alpha$ lies in the range between two angles $a$ and $b$, that span, for example, a prograde orbit configuration, as 
\begin{equation}
 p_\mathrm{pro}(a,b) := \int_{a}^{b} p(\alpha|D) \;\mathrm{d}\alpha\,.\label{eq:prob}
\end{equation}
The probability density function $p(\alpha|D)$ is the \textit{posterior} probability distribution, holding the probability that $\alpha$ is an angle leading to a prograde (or retrograde) orbit, given the data $D$.
According to \textit{Bayes' theorem} we define the posterior as
\begin{equation}
 p(\alpha|D) = \frac{p(\alpha)\,\mathcal{L}(D|\alpha)}{\int_{0}^{2\piup} p(\alpha)\,\mathcal{L}(D|\alpha) \;\mathrm{d}\alpha}\,,
\end{equation}
where $p(\alpha)$ is the \textit{prior} probability distribution that any $\alpha$ is given, which is $(2\piup)^{-1}$ for a uniform distribution, and $\mathcal{L}(D|\alpha)$ is the \textit{likelihood} for the
data $D$ given some value of $\alpha$.
The \textit{Bayesian evidence} given in the denominator holds both distributions as well.
We further define the likelihood as
\begin{align}
 \mathcal{L}(D|\alpha) &= \prod_j^N \frac{1}{\sqrt{2\piup\sigma_j^2}}\cdot \mathrm{e}^{-\frac{(d_j-m_j(\alpha))^2}{2\sigma_j^2}}\\
 &= C \cdot \mathrm{e}^{-\frac{1}{2}\chi_0^2}\cdot\mathrm{e}^{-\frac{1}{2}(\chi^2(\alpha)-\chi_0^2)}\,,
\end{align}
under the assumption of Gaussian errors and with $\chi_0^2$ as an offset, for example, of the null-hypothesis $\alpha =0\degr$.
These two constant factors are the same for the Bayesian evidence, and therefore vanish in the posterior distribution.
We therefore find the posterior as
\begin{equation}
 p(\alpha|D) = \frac{\mathrm{e}^{-\frac{1}{2}\Delta\chi^2(\alpha)}}{\int_0^{2\piup} \mathrm{e}^{-\frac{1}{2}\Delta\chi^2(\alpha)}\;\mathrm{d}\alpha}\,.
\end{equation}
In this framework we are able to distinguish between prograde and retrograde orbit configurations from multiple transit light curves alone by integrating the posterior between the corresponding limiting angles $a$ and $b$. 

\subsection{Observational data: The Kepler-20 system}

To apply our method of analysis we need a data set of a multi-transiting system with high photometric quality, low stellar activity, a time resolution covering possible PPOs, and 
of sufficient duration to contain a few observed multiple transits.
Among the numerous multiple planet systems discovered by the \textit{Kepler} satellite, we decided to use the Kepler-20 system \citep{Borucki2011} for our investigations.
This solar-like star hosts six planets, five of which perform transits \citep{Buchhave2016}.
The known radii of the transiting planets are in the range of Earth to super-Earth size, where Kepler-20\,e even tends to be a sub-Earth (see Table~\ref{tab:app_paras}).
No Rossiter-McLaughlin measurements of the system are available because of the low RV signals induced by the small planets \citep[see][]{Buchhave2016}, and no other alignment measurements exist.
Additionally, as already presented by \citet{Gautier2012} and later also by \citet{Buchhave2016}, none of the transiting planets shows significant TTVs and their orbits are almost circular.
The stellar activity of Kepler-20 is quite similar to that of our Sun in its active state in the sense of period and variability, however Kepler-20 has a slightly larger peak-to-peak amplitude \citep{Gautier2012}.
All these properties make Kepler-20 an ideal target to test our technique, where all other methods to measure the alignment between the orbits or between the orbits and the stellar spin axis have failed so far.

We downloaded all available quarters of Kepler-20 (Q1-Q17) from the NASA Exoplanet Archive\footnote{\url{exoplanetarchive.ipac.caltech.edu/}}.
We used the short-cadence PDCSAP flux and ended up with 1331 days of Kepler-20 observations (Q3-Q17).
For consistency, we carried out transit fits of all transiting planets in the system.
First we normalized all transit light curves by a second-order polynomial followed by a fit of the individual orbital periods. The transit fits of all planets were then carried out 
with the above-mentioned \texttt{occultquad} routine  using fixed limb-darkening coefficients (LDCs; see Sect.~\ref{subsec:LD}). The results were used as initial values for a 
Markov-chain Monte-Carlo (MCMC) sampling of the parameters to acquire reliable error estimates. 

In Table~\ref{tab:k20_multi}, we summarize the available multiple transits of the Kepler-20 system.
These are determined by counting model multi-transits of the form given in Eq.~\ref{eq:multi_mod}, which are deeper than the corresponding individual transits.
\begin{table}
\caption[]{Number of unique multiple transits}
\label{tab:k20_multi}
\centering
\begin{tabular}{cr|cr}
\hline\hline
Planets & Number & Planets & Number \\
\hline
b+c & 7 & d+e & 1\\
b+e & 11 & d+f & 1\\
b+f & 5 & e+f & 2\\
c+e & 3 & c+e+f & 1\\
c+f & 1 & & \\
\hline
\end{tabular}
 \tnotecaption{Based on Kepler-20 short-cadence data.}
\end{table}
In total, we find 32 multiple transits observed by Kepler, of which 23 occur with planet (b), and 18 with planet (e), out of which one is a triple-transit.
Both planets show the shortest orbital periods in the system, namely $3.7$ and $6.1$ days, increasing the probability of showing multiple transits with one of them.
As planets (e) and (f) are almost half as large as planet (b), providing a transit depth-to-noise ratio, $\delta/N$, of only 0.19 and 0.24, we limit our investigations 
to the seven multiple transits of planets (b) and (c) to derive the most significant results.

\subsection{Stellar limb darkening}
\label{subsec:LD}
Stellar limb darkening is known to be important for transit modeling. There are different ways to treat this parameter in transit studies. For example, it can be considered 
as a fixed input parameter by relying on limb darkening calculated from model atmospheres \citep[e.g.,][]{Claret2012}, or, with sufficient photometric precision,
the LDCs can actually be measured \citep{HM2013}.
For the Kepler-20 system, we assume that it is appropriate to use one 
limb-darkening model for all transits in the Kepler-20 system, while  \citet{Buchhave2016} used 
fitted stellar quadratic LDCs obtained by transit modeling of the individual planets.
The derived LDCs differ significantly from planet to planet, which we consider unphysical.
Because of the strong correlation between limb darkening and the other transit parameters \citep[e.g.,][]{HM2013}, we insist on using one 
limb-darkening description for a given star, which should lead to more reliable results. 

For this work, we decided to fix the LDCs ---using the quadratic law--- to values obtained by a fit to the intensity distribution of a recent 
PHOENIX model atmosphere\footnote{PHOENIX 18.3, P. Hauschildt private communication.} generated using the stellar parameters of Kepler-20, namely $T_\mathrm{eff} = 5495$\,K, $\log g=4.446$, and $[\mathrm{m}/\mathrm{H}]=0.07$ as given in \citet{Buchhave2016}. 
We integrated the PHOENIX intensities in the Kepler passband \citep{kHandbook} and performed a least-squares fit to the intensity distribution leading to $u_1=0.495$ and $u_2=0.154$.
We note that these quadratic LDCs were only used in our transit parameter fits, while we use a different approach for our obliquity sampling, which is described in Sect.~\ref{sec:blender_setup}.

\subsection{Star-spot modeling}\label{sec:spot_map}
To obtain a general idea of the phenomenology of the light curve and to potentially identify active longitudes on the host star, we carry out an activity analysis.
First we determine the stellar rotation period using a generalized Lomb-Scargle periodogram of the data \citep{Zechmeister2009}, and
second, we construct a brightness map from the light curve.
We do this by splitting the data into chunks with a duration equal to the stellar rotation period. 
The measured flux is then color-coded and we plot the chunks with respect to time (y-axis) and rotational phase/longitude (x-axis) to construct this map.
Because we assume the photometric variations to be induced by spots, we carry out a spot inversion technique to investigate the spot distribution on Kepler-20.
We use a spot model similar to the one used by \cite{Ioannidis2016}, including a reasonable temperature contrast,
which allows us to determine spot longitudes and sizes that are interesting for our analysis.
However, because of the correlation between spot temperature, spot latitude, and size, our results can only serve as a lower limit on the spot size.

\subsection{Modeling multiple transits: Using Blender for light-curve synthesis}
There are several algorithms able to simulate PPOs, such as \citet{Pal2012},
yet we decided to choose an entirely different approach to synthesize multiple transits including PPOs.
For this purpose, we use the publicly available 3D rendering and animation software Blender\footnote{\url{blender.org}. 
We note in passing that this software is not to be confused with the \texttt{BLENDER} algorithm \citep{Torres2004, Torres2011} 
used to measure and estimate blends in aperture photometry of the \textit{Kepler} satellite.}.
This software has already been used in astrophysical contexts, for example by  \citet{Naiman2016} and \citet{Garate2017} for visualization purposes;
however, as far as we are aware, we are the first to use Blender for exoplanet transit light-curve synthesis.  

\subsubsection{Advantages of Blender}
\label{sec:blender}
Blender uses a physical ray-tracing engine called \textit{Cycles}, which is used for very realistic 3D animations, and can in our case be used to simulate planetary systems.
This includes, in particular, the simulation of diffuse Lambertian reflection caused by object surfaces, such as those of spheres (which we use as planets).
These spheres can then be put in an orbit around a larger, light-emitting sphere as is the case in a real planetary system.

There are some advantages of using Blender for transit light-curve synthesis in comparison to other codes capable of simulating PPOs.
First, stellar activity caused by spots and faculae can easily be added to the surface of the simulated star, for example by using a texture or 
by directly editing the surface vertices.
This can also include a time dependence, by adding a simple rotation to the star-like object in the scene.
However, this requires a reasonable spot and faculae model to precisely describe the global light-curve variations, 
which is not a trivial task in itself \citep[e.g.,][]{Huber2010}.
Thanks to Blender's 3D object modeling capabilities, it is self-evident that all involved objects can have arbitrary shapes, for example ellipsoids representing stellar and planetary oblateness or even
rings.
As Blender can be fully controlled via Python scripts and has its own integrated Python console, all the settings and simulations can be automated and are reproducible.
Furthermore, and most importantly, it is possible to include stellar limb darkening obtained from model atmospheres directly, excluding systematic uncertainties introduced by fitting conventional limb-darkening laws to the radiation field of the 
model atmosphere, an outstanding capability in combination with the other advantages stated above.
Even without PPOs, codes including non-analytical limb darkening are rare and require numerical transit simulation algorithms such as that introduced by \citet{myThesis}.
This makes Blender  an alternative to these codes, as it is capable of simulating light curves in arbitrary wavelength bands for any instrument and filter combination.

Nevertheless, the necessary ray tracing requires a considerable amount of computational power, the \textit{Cycles} engine can run the simulations on graphic cards (GPUs) that are 
optimized for such calculations by utilizing the NVIDIA CUDA or AMD OpenCL frameworks. 
In this way, the simulations are carried out as fast as possible.
Before we proceeded with our Blender simulations, we confirmed the consistency between exact calculations and Blender calculations by demonstrating 
that the residuals between a transit generated with the \texttt{occultquad} routine and Blender are of the order of $10^{-7}$ when identical limb-darkening laws are used.
This is about three magnitudes lower than the standard deviation of the used data set, and therefore we are confident that Blender yields accurate results. 
Furthermore, and relevant in the context of our work, we reproduce the prototype of PPOs \citep{Hirano2012} with Blender.
We use the orbit and planet parameters given by \citet{Masuda_2013} as input and let Blender simulate the multiple transit presented in these works.
The result is shown in Fig.~\ref{fig:masuda}, including a PPO event that nicely reproduces the observed data.

For more detailed information about Blender, we refer to the official manual\footnote{\url{blender.org/manual/}}, which describes the functionality of the software in detail, and to \citet{Garate2017} for more references.
In the following section, we describe our Blender setup to generate the simulated transit light curves of the Kepler-20 system.

\subsubsection{Blender setup}
\label{sec:blender_setup}
 We used Blender $2.79$ for our simulations, controlling all settings with a Python script.
We note that all the settings can also be made in the graphical user interface (GUI), which was actually our first approach to understand how Blender works.
Each parameter in the GUI shows the corresponding Python variable in a mouse-over text which helps to create the automated script. 

To set up Blender for our simulation purposes, we use the following steps:
\begin{itemize}
 \item[1.] Set up the Blender scene: We have to delete all preset objects present after Blender has started.
 We then add the objects needed for the simulation, that is, six spheres for the Kepler-20 system, one star, and five planets.
 Thereby the star has a normalized radius of 1 and is placed at the origin of the three-dimensional Cartesian coordinate system.
 The sizes and distances of the planetary spheres are set according to the corresponding fit results of $p$ and $a/R_\mathrm{S}$.
 We choose the Blender mesh object ``UV sphere'' for the objects, and set the number of segments and rings to 256 in order to obtain smooth shapes.
 To achieve a light-emitting star, one has to add a surface material that is set to emission with strength equal to $1.0$.
 The color should be pure white.
 
 \item[2.] Orbital motion: The usage of circle-curve objects around the star-like sphere as animation paths is not discussed here, although it provides the opportunity to view and edit the animation in the GUI directly.
 We prefer to calculate the x-y-z-positions of the $j$-th planet for each time-step $n_f$ according to the relations
 \begin{align}
  x_j &= -1 \cdot a_j \cdot \sin(\varphi_{0,j} + \omega_j\, n_f) ,\\
  y_j &= -1 \cdot a_j \cdot \cos(\varphi_{0,j} + \omega_j\, n_f) + \cos(i_j),\\
  z_j &= -1 \cdot a_j \cdot \cos(\varphi_{0,j} + \omega_j\, n_f) + \sin(i_j)\,.
 \end{align}
 These relations are valid for circular orbits, with $\varphi_{0,j}$ being the initial phase shift of the corresponding planet,
 which is calculated using the fit results of the times of the first transit centers and the orbital periods according to 
 \begin{equation}
  \varphi_{0,j} = \left( \frac{-T_{0,j}}{\Delta t} \right) \omega_j\,,
 \end{equation}
 where the angular velocity $\omega_j$ can be written as
 \begin{equation}
  \omega_j = \left( \frac{\Delta t}{P_{\mathrm{Orb},j}} \right) \cdot 2\piup\,.
 \end{equation}
 The time $\Delta t$ between two simulated frames $n_f$ is set to 60 seconds to match the time-resolution of the Kepler-20 short-cadence data. 
 The total number of frames to be simulated is limited to the duration of the seven multiple transits of planets b and c.
 Additionally, we add $0.05$ days as out-of-transit parts to both sides of the transits, leading to 2944 frames to be rendered for one set of transits.
 
 The tilt angle $\alpha$ varied for our obliquity sampling is induced by a rotation matrix of the form
 \begin{equation}
  \textbf{R}_y(\alpha) = \left( \begin{array}{ccc}
                        \cos(\alpha) & 0 & \sin(\alpha)\\
                        0 & 1 & 0 \\
                        -\sin(\alpha) & 0 & \cos(\alpha)\\
                       \end{array}\right)\, ,
 \end{equation}
 which rotates the orbit of planet b around the y-axis.
 This leads to a new position vector of planet b given by
 \begin{equation}
  \left(
  \begin{array}{c}
          x_\mathrm{b}'\\
          y_\mathrm{b}'\\
          z_\mathrm{b}'\\
  \end{array}
  \right) = 
  \left(
  \begin{array}{c}
        x_\mathrm{b}\cos(\alpha) + z_\mathrm{b}\sin(\alpha) \\
        y_\mathrm{b} \\
        -x_\mathrm{b}\sin(\alpha) + z_\mathrm{b}\cos(\alpha)\\
  \end{array}
  \right)\,.
 \end{equation}
We vary the tilt angle in $1\degr$ steps after all seven transits have been simulated.
According to Eq.~\ref{eq:a_start} to Eq.~\ref{eq:end}, 294 angles are needed to complete our model grid. 
The only frames to be rendered are those where both planets are in front of the stellar disk at the same time.
Thus, there is no need to repeat the out-of-transit simulation for every angle, reducing the number of frames to 715 per tilt angle,
resulting in 212\,439 frames in total.

 \item[3.] Stellar limb darkening: The star-like sphere is treated as an isotropic light emitter, that is, it is identical to a uniform source in transit modeling.  
 The simulation would now already lead to transits without limb darkening, which are well suited to test the whole setup.
 A simple way to include limb darkening of the host star is to add a plane object, which acts like a filter placed in the line-of-sight.
 We add a texture to this plane, which is transparent according to its pixel values.
 The texture is a $30\,001^2$\,px image of a stellar disk, which is generated using the limb-intensity distribution from the PHOENIX atmosphere used for Kepler-20. 
 We interpolated linearly between the disk-position intensities predicted by PHOENIX to achieve this almost 1\,Gpx image, which is needed to achieve optimum results.
 We then proceed without adding spots or faculae to the stellar disk brightness distribution.
 Also, Kepler-20 is a slow rotator comparable to the Sun, and we neither include gravity darkening nor oblateness.

 \item[4.] Render and light-curve acquisition:
 For a rendered image one needs a ``camera'' to be placed in the scene.
 For Kepler-20 we set its position to $(0,-80,0)$, where the value is chosen according to the largest semi-major axis in the system.  
 It is important that the camera lens is set to ``orthographic mode'' to achieve parallel light rays, as is the case for an observer at an infinite distance.
 The orthographic scale has to be $\geq 2$ to fit the star inside the viewing window.
  Before the rendering is started one needs to set the render engine to ``Cycles Render'' manually, and the ``Color Management'' has to be set to ``Raw''.
 
 We choose a resolution of 500x500\,px for our simulations, which is sufficient for our purposes.
 In contrast to visualizations, for which Blender is normally used, a scientific setup leads to images that are mostly black.
 Only the star-like light-emitting sphere is visible, while the planet remains invisible to the naked eye until the eclipse starts.
 These images are saved as a 16-bit gray-scale lossless-compressed png file format.
 A flux value is obtained by integrating the pixel values of this image, which can be normalized be the integral of an image representing the out-of-transit time-step.
 In this way, a light curve is computed frame by frame.
 With the total amount of frames rendered, this takes almost 50\,GB of memory.
\end{itemize}
It is important to note that Blender applies a Fourier-Smoothing to object edges using the \textit{Blackman-Harris} pixel filter, resulting in much more realistic scenes 
and reducing aliases visible at sharp edges which do not lie parallel to pixel rows or lines in the final image.
For moving objects, like planets, this results in a planetary disk which always
shows the same size, but with some gray pixels at the transition 
from the black disk to the white background (star).
This also applies for the limb-darkening filter plane at the transition from the stellar limb to the black background.
Thus, the resulting brightness distribution over the stellar disk shows slight differences when compared to the input limb darkening.
To remove this effect, one has to set the pixel filter algorithm to ``Box''. 
With the setup mentioned above, the machine we used, which is equipped with two Titan Xp GPUs, renders 2.42 frames per second.
The whole model grid for one case, as introduced in Sect.~\ref{sec:oom}, takes $24.4$ hours of rendering time.

\section{Results}
\subsection{Transit modeling}
The results of our transit modeling of Kepler-20\,b and c are listed in Table~\ref{tab:paras}, while we give the results for planets d, e, and f in Table~\ref{tab:app_paras} in the Appendix.
For a better comparison of our results, we also include the values for planets b and c as derived by  \citet{Buchhave2016} in Table~\ref{tab:paras}.
Our fit results represent the parameter sets determined from the MCMC parameter traces as the lowest deviance solution, which maximizes the likelihood.
The $1\sigma$ errors of our fit results represent the 68.3\,\% credibility intervals of the posterior parameter distributions. 
\begin{table*}
\renewcommand{\arraystretch}{1.25}
\caption[]{Transit parameters of Kepler-20\,b and c}
\label{tab:paras}
\centering
\begin{tabular}{lccccc}
\hline\hline
Parameter & \citet{Buchhave2016} & This Work & \citet{Buchhave2016} & This Work\\
\hline
&\object{Kepler-20\,b}& & \object{Kepler-20\,c} &\\
\hline
$R_\mathrm{P}\,/\,R_\mathrm{S}$ & $0.01774^{+0.00053}_{-0.00003}$      & $0.01805^{+0.00013}_{-0.00018}$          & $0.02895^{+0.00029}_{-0.00006}$ &$0.02878^{+0.00001}_{-0.00001}$\\
$i\,/\,$\degr             & $87.355^{+0.215}_{-1.594}$           & $87.687^{+0.566}_{-0.365}$               & $89.815^{+0.036}_{-0.632}$ & $89.967^{+0.053}_{-0.124}$\\
$a\,/\,R_\mathrm{S}$            & $10.34^{+0.20}_{-0.32}$              & $10.60^{+0.44}_{-0.31}$                  & $21.17^{+0.59}_{-0.51}$& $22.764^{+0.004}_{-0.020}$\\
$T_0\,/\,$d$^\mathrm{a)}$      & $967.502014^{+0.000253}_{-0.000217}$  & $967.501962^{+0.000058}_{-0.000038}$     & $971.607955^{+0.000248}_{-0.000202}$& $971.608470^{+0.000037}_{-0.000051}$\\
$P_\mathrm{Orb}\,/\,$d & $3.69611525^{+0.00000115}_{-0.00000087}$      & $3.69611590^{+0.00000015}_{-0.00000027}$ & $10.85409089^{+0.00000303}_{-0.00000260}$ & $10.85408479^{+0.00000078}_{-0.00000029}$\\
$u_1$$^\mathrm{b)}$                           & $0.427^{+0.120}_{-0.051}$            & 0.495 & $0.393^{+0.060}_{-0.038}$& 0.495\\
$u_2$$^\mathrm{b)}$                           & $0.295^{+0.134}_{-0.078}$            & 0.154 & $0.408^{+0.052}_{-0.069}$& 0.154\\
\hline
\end{tabular}
\tnotecaption{$^\mathrm{a)}$\, The time of the first transit center is given in BJD$-2.454\cdot10^6$\,d.
$^\mathrm{b)}$\,Quadratic LDCs used in this work are fixed to values obtained by a fit to model intensities (Sect.~\ref{subsec:LD}).   }
\end{table*}

We find that most of our MCMC parameter results do not differ significantly from those given by \citet{Buchhave2016}.
The orbital periods and the times of the first transit centers show the lowest deviations of the parameters and are almost identical.
In the case of planet b, we determine a difference for $T_0$ of only $4.5$\,s, which is four to five times smaller than the error.
For planet c, the situation is slightly different: our $T_0$ appears almost one data point later and $P_\mathrm{Orb}$ is $0.5$\,s shorter. 
Both differences are relatively small but significant beyond the given $1\sigma$ uncertainties.
The remaining transit parameters of planet b are all consistent with previous results, while our results for planet c tend to show a slightly smaller planet 
and a larger semi-major axis.
In the case of the radius ratio, this amounts to only $0.5$\,\%,  and to an approximately 7$\,\%$ greater distance to its host star, both significant when considering the $1\sigma$ uncertainties.
Overall, it appears that our results show smaller uncertainties when compared to the previous results.
When considering only the multi-transits of planet b and c, which are important for our work, we find a lower value of the reduced $\chi^2$ when using our fit results for the models, namely $\chi^2/\nu = 1.052$ in contrast to $\chi^2/\nu = 1.115$, when the literature fit results are adopted.
Both results, ours and those from \citet{Buchhave2016}, are used for our obliquity sampling approach to investigate possible differences in the alignment results.

\subsection{Stellar activity}
\label{sec:res_spots}
As introduced in Sect.~\ref{sec:spot_map} we performed an activity analysis of the data.
The light curve of Kepler-20 shows variability with a peak-to-peak value of about $0.1\,\%$ to $0.2\,\%$, while we find the largest variation of $0.6\,\%$ in Q17.
We interpret the modulations as star spots becoming visible as the star rotates.
We find a solar-like rotation period of $26.948\,$d\,$\pm\,0.001\,$d, which is in good agreement with the previous results of \citet{Gautier2012} and \citet{Buchhave2016}.

\begin{figure}
\centering
 \includegraphics[width=0.5\textwidth]{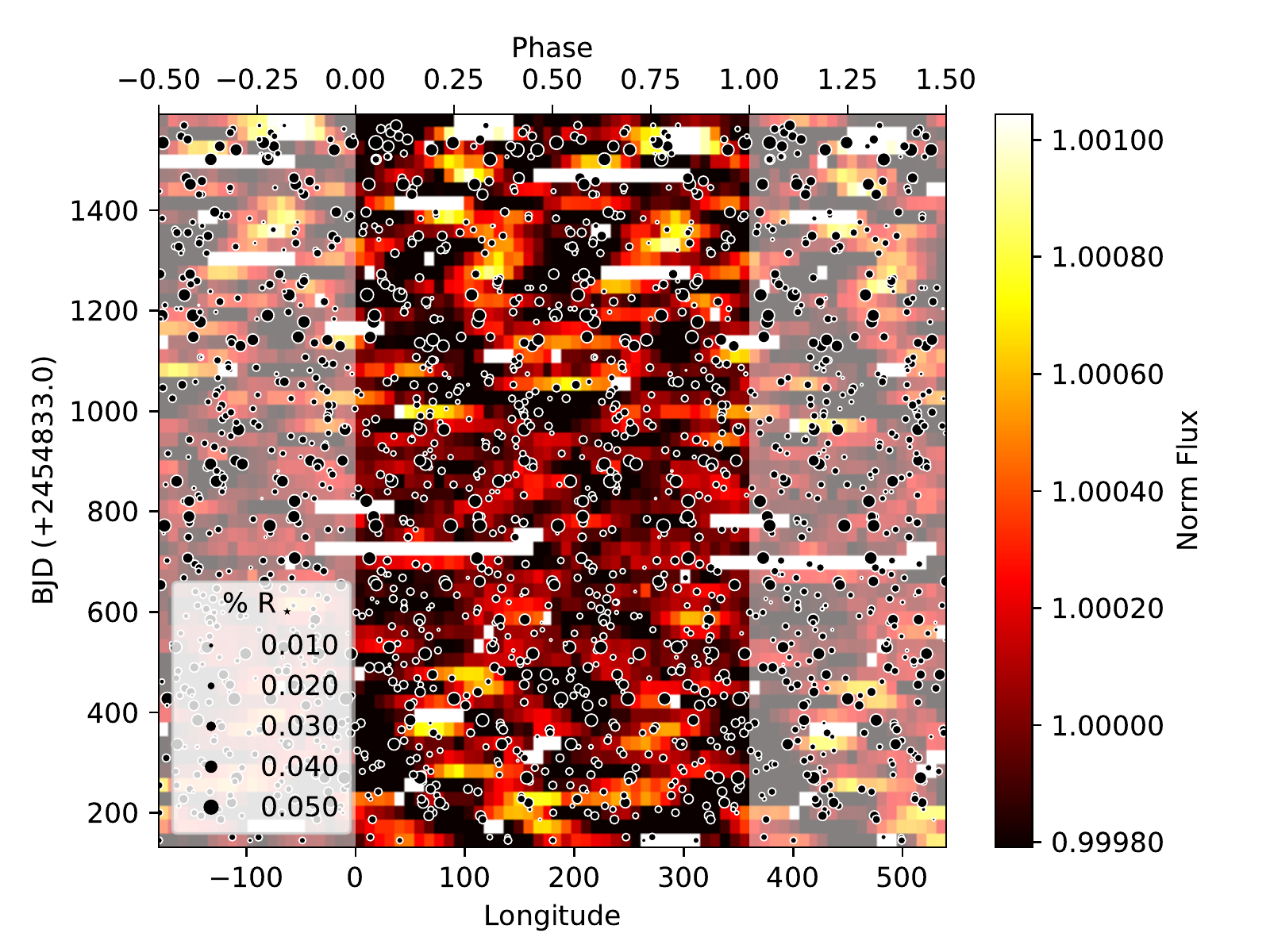}
\caption{Stellar activity map of Kepler-20.
Relative fluxes are color coded.
\textit{Black dots} show the positions and sizes of the spots.
\textit{Shaded areas} indicate the repeated parts of the light curve (see text for details). }
\label{fig:actmap}
\end{figure}

In Fig.~\ref{fig:actmap} we show the brightness map of Kepler-20. 
The color code shows the fluctuations in time (y-axis) and longitude (x-axis), where bright regions show higher relative flux and darker regions lower relative flux, respectively. 
To assist the viewer's perspective of the light curve and so as not to interrupt the modulations, we plot the second half of each light-curve chunk before its subsequent chunk (phase $<0$) and the first half of the subsequent chunk at phase $>1$ as shaded areas.

In agreement with the phenomenology of the light curve, we do not distinguish any long-term active regions such as those visible as dominant and long-lived brightness oscillations on very active stars.
Here, the signal is not clearly periodic and there is no evidence for active longitudes as seen on other stars (see, \citet{Ioannidis2016} and \citet{Ioannidis2020}).
This is supported by the results of our spot inversion, which are shown as dots in Fig.~\ref{fig:actmap}.
The diameter of each dot represents the predicted size of
those spots; we note again that the spot size is strongly correlated with the spot temperature
as well as the latitude of the spot, and therefore we can only measure a minimum value for
the spot radius.

Our analysis shows that we cannot identify significant spot-crossing events within the \textit{Kepler} data ---which reappear phase-shifted after several orbital periods--- either for planet b or for planet c.
This could have indicated a projected spin--orbit angle of $\lambda$ close to 0\degr~similar to that reported for Kepler-17 by \citet{Desert2011}.
Therefore, there is no prior evidence for any spin--orbit alignment or mutual orbital alignment in the data that could help us to interpret the results from our orbital obliquity sampling.

\subsection{Obliquity sampling results}

\subsubsection{Simulation setup}
\begin{figure*}
   \centering
   \includegraphics[width=0.98\textwidth]{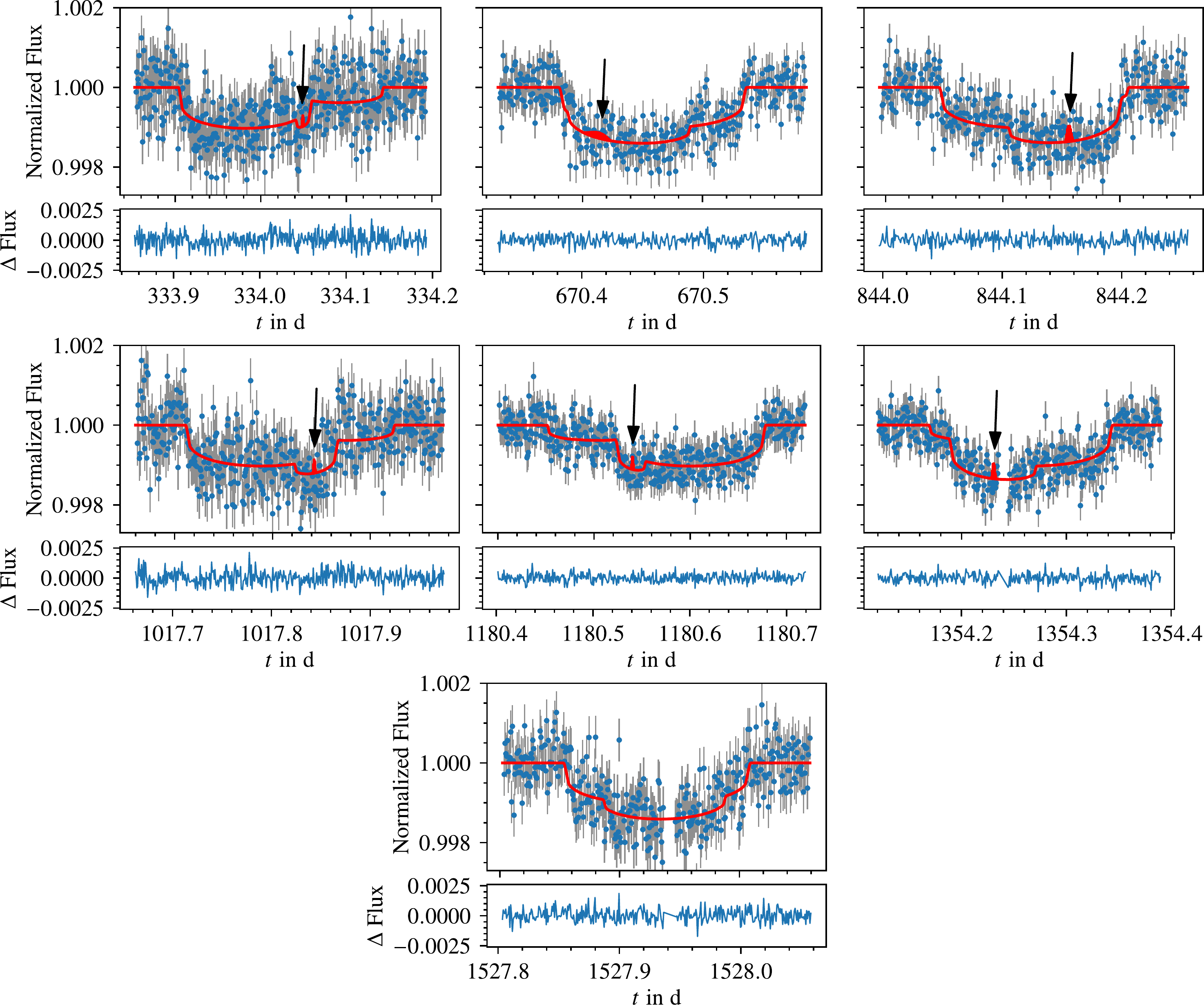}
   \caption{Multiple transits of Kepler-20\,b + c (\textit{dots}) together with our Blender models (\textit{red}).
   The error bars indicate the $1\sigma$ error of the data taken from the FITS files.
   Simulated PPOs present in six out of seven transits are indicated by \textit{arrows}.
   We plot every Blender model resulting from our set of $\alpha$ values on top of each other in order to visualize the variation in position of the PPOs in the transit. 
   Below every transit, we show the residuals between the data and our Blender models for the null-hypothesis $\alpha = 0$\textdegree.}
   \label{fig:transits_ppo}
\end{figure*}

As outlined in Sect.~\ref{sec:blender_setup}, we apply our obliquity sampling to the Kepler-20 planets b and c.
Using the Blender software we simulate the seven multiple transits shown in Fig.~\ref{fig:transits_ppo}.
For that, we use two different sets of transit parameters for the planets, namely the best-fit values of the transits from our and previous work (Table~\ref{tab:paras}).
The precise start and end times of every multiple transit result from model transits of the form given in Eq.~\ref{eq:multi_mod} using the mentioned parameter sets.
The mid-times of the multi-transits are $334.02$\,d, $670.45$\,d, $844.13$\,d, $1017.82$\,d, $1180.56$\,d, $1354.25$\,d, and $1527.93$\,d.
The model transits are evaluated at the Kepler-20 observation times, leading to about one data point per minute.
The model multi-transits are stored as ASCII files with time and normalized flux.

We compute the model light curves for various values of the orbital tilt angle $\alpha$.
According to Eqs.~\ref{eq:a_start} to \ref{eq:end}, these are for example $20\degr$ to $158\degr$, and $202\degr$ to $340\degr$, in increments of $1\degr$. 
However, as mentioned above, we start at $\alpha = 0\degr$.
We store the model light curves for each angle separately and repeat the process for cases 1 and 4 (see Fig.~\ref{fig:cases}) for both parameter sets.
In total, we produce a model grid that consists of about $8.5 \cdot 10^5$ data points.

The simulations we carried out were able to generate artificial PPOs in the Kepler-20 system for some specific orbital tilt angles $\alpha$.
These PPOs appear as small bumps in six out of seven multiple transits of planets b and c indicated by arrows in Fig.~\ref{fig:transits_ppo}.
These artificial PPOs differ from transit to transit in size and position, and also do not appear for the same angle $\alpha$. 
However, this is not to be expected because of the different phase angles of planets b and c in each multiple transit, indicating incommensurable orbital periods.
In general, we note that there is a minimum angle per transit where the planets begin to virtually overlap and one maximum angle beyond which PPOs no longer occur.
Between these angles the PPOs slightly shift in time depending on $\alpha$.

\subsubsection{Goodness of fit as a function of $\alpha$}
To assess the goodness of fit between data and model, we compute the $\chi^2$-statistics for every chosen value of $\alpha$.
We do this by treating all seven multiple transits at once as one data set, as done for transit parameter modeling.
We use the standard error of the Kepler-20 data points given in the FITS files to compute the $\chi^2$.

\begin{figure*}
   \centering
   \includegraphics[width=0.48\textwidth]{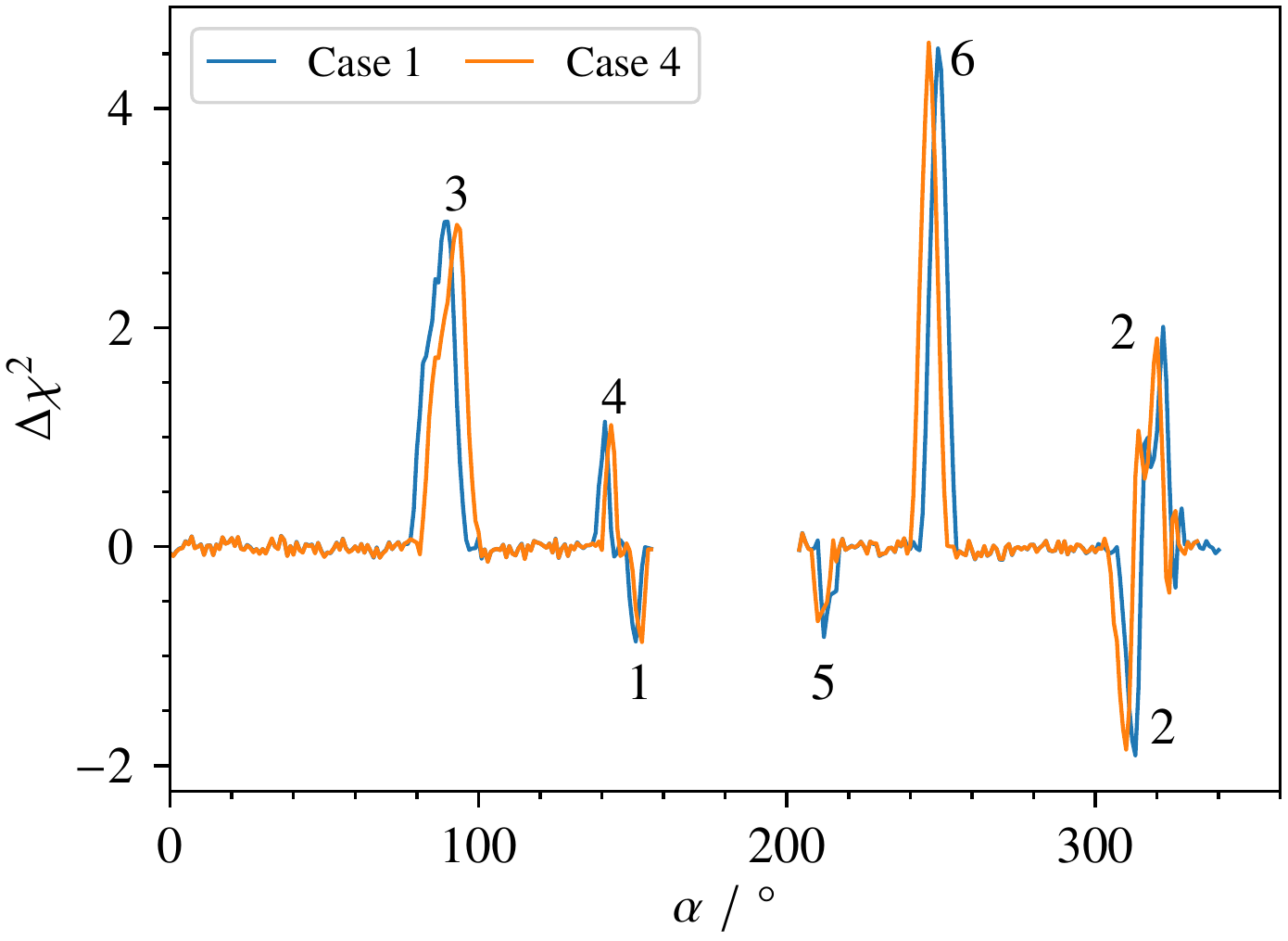}
   \hfill
   \includegraphics[width=0.48\textwidth]{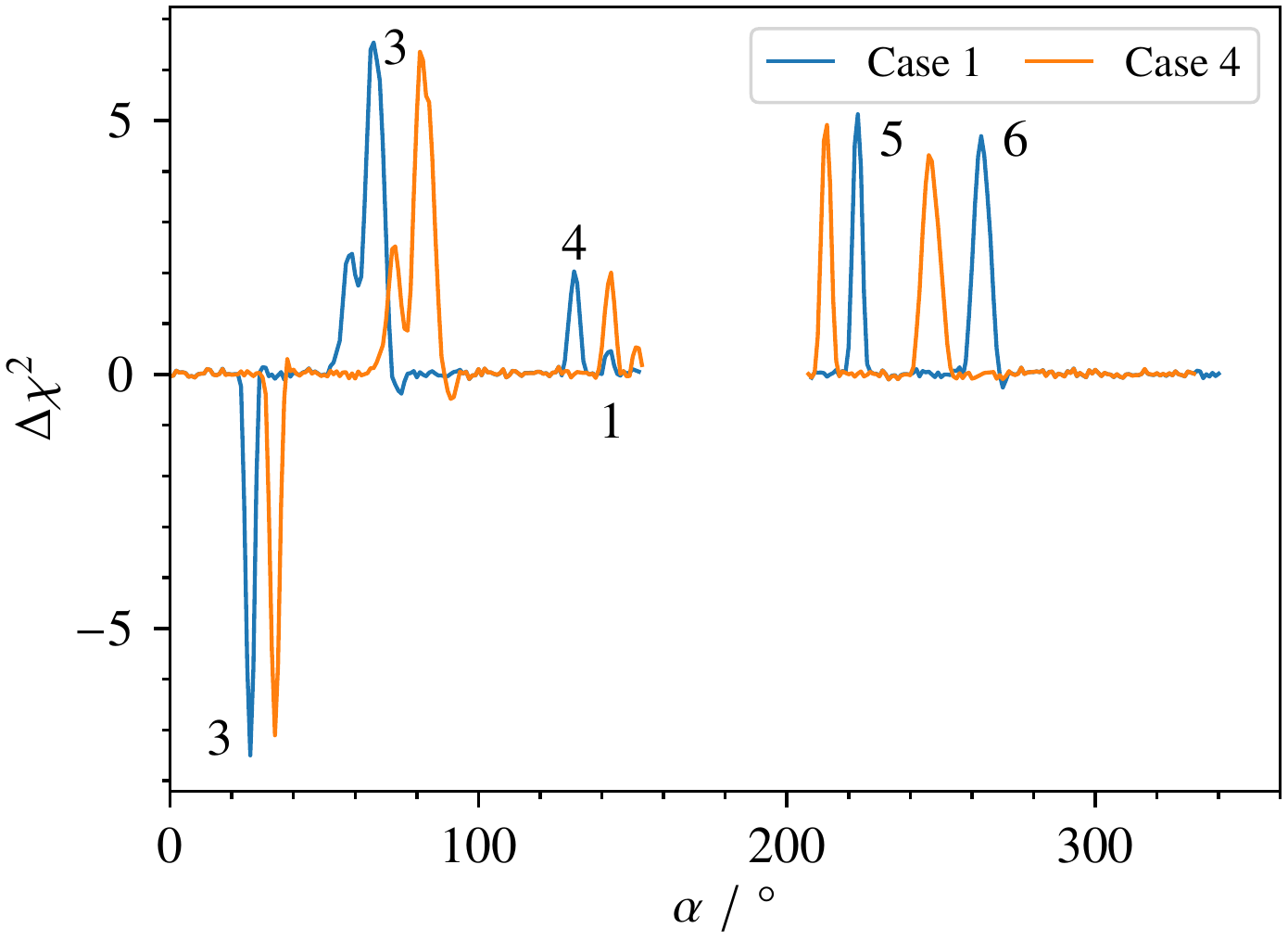}
   \caption{Change of the $\chi^2$ value between model and data as a function of the orbital tilt angle $\alpha$.
   The \textit{left} panel shows the results for simulations carried out with our transit parameter fits, while the \textit{right} panel holds results when transit parameters from \citet{Buchhave2016} are used. 
   The two simulated cases are shown in different colors (\textit{labels}).
   The numbers indicate the transit in which artificial PPOs lead to $\Delta\chi^2 \neq 0$.}
   \label{fig:delta_chi_sq}
\end{figure*}

The presence of artificial PPOs for some angles leads to a change in the $\chi^2$ value calculated between the model and the Kepler-20 multiple transits.
In Fig.~\ref{fig:delta_chi_sq} we present this {change} of $\chi^2$ as a function of $\alpha$ for our cases 1 and 4 based on our transit parameter fits and those of \citet{Buchhave2016}.
To achieve this $\Delta\chi^2$, we subtracted the $\chi^2$ value of the null hypothesis, that is $\alpha=0$\textdegree~and no PPO, or where the artificial transits are not different from that.
For our transit parameters, we find $\chi^2_\mathrm{M}= 2980$, while the previous transit parameters lead to $\chi^2_\mathrm{B}= 3160$, which is a significantly higher value.

As visible in Fig.~\ref{fig:delta_chi_sq}, the ground-level of the $\chi^2$ value shows a noise-like behavior with a standard deviation of $\pm0.05$, although it should theoretically show a constant value for angles where no artificial PPOs are present.
This is a reproducible effect induced by Blender, which can be reduced by increasing the resulting image resolution or the number of rendered samples, both having a negative impact on the computational speed.
We therefore attribute this effect to numerical noise.
For the determination of the mentioned values of $\chi^2_\mathrm{M}$ and $\chi^2_\mathrm{B}$ we used the median value of the first 20 angles, where  no PPOs occur for either of the transit parameter sets.

\subsubsection{Properties of the $\Delta\chi^2$ curves}

Both results show a systematic shift of the peaks and dips when the different cases are compared.
Below 180\textdegree~, case 4 shifts to larger angles, while above 180\textdegree~the opposite occurs.
This shift is caused by the fact that both planets show a nonzero impact parameter.
As illustrated in Fig.~\ref{fig:cases}, cases 1 and 4 differ in the sense that both planets are either on the same hemisphere or are not.
If one planet were to show an impact parameter of zero, there would not be any difference.
This effect is stronger for the results obtained using the previous transit parameter fits, as this parameter set shows larger impact parameters compared to ours.

It is clearly visible in Fig.~\ref{fig:delta_chi_sq} that the $\Delta\chi^2$ curves obtained from the different planetary parameter sets differ significantly from each other.
While the right panel shows only one very prominent dip at $\alpha=26$\textdegree~, we identify about four dips in the left panel that are less significant.
Both results share the characteristic that they show about five peaks per case where the $\chi^2$ is increased.
Additionally, the order in which angle PPOs occur and in which transit (3, 4, 1, 5, or 6) is the same because the phase angles of the multiple transits do not differ significantly between the models.
Transit number 7 shows no PPOs in both cases, while transit number 2 is prominent only for our results.
Although the four dips in the left panel represent a less significant change of the model (given the data) than the deepest dip in the right panel, it should be borne in mind that in the sense of a $\chi^2$ statistic our fit results of the planetary parameters used for the simulations represent the data significantly better than the previous results leading to the right panel.
Therefore, we argue that even the model leading to the dip in the right panel provides a poorer representation of the data than our model for the null hypothesis. 

\subsubsection{Interpretation of the results}

The question now arises as to how we can interpret this $\Delta\chi^2$ curve.
In general, we note that a higher value of $\chi^2$ indicates a model that is less consistent with the data.
If we recall that the only parameter varied in the simulations is the tilt angle $\alpha$, we can interpret angles that lead to a positive $\Delta\chi^2$ as being inconsistent with the data and not favorable for the system's geometry.
Inversely, this means that angles that lead to negative $\Delta\chi^2$ values lead to a model that better describes the data.
We list the angles of the maximum and minimum of the peaks and dips in Table~\ref{tab:angles}.
\begin{table}
\renewcommand{\arraystretch}{1.25}
\caption[]{Orbit angles of Kepler-20\,b leading to artificial PPOs.}
\label{tab:angles}
\centering
\begin{tabular}{rrr|rrr}
\hline\hline
$\alpha_1\,/\,\degr$ & $\alpha_3\,/\,\degr$ & $\Delta\chi^2$ &
$\alpha_4\,/\,\degr$ & $\alpha_2\,/\,\degr$ & $\Delta\chi^2$ \\
\hline
$90^{+5}_{-11}$ & $270^{+11}_{-5}$ & $2.97$ & $93^{+6}_{-11}$ & $267^{+11}_{-6}$  & $2.94$ \\
$141^{+1}_{-2}$ & $219^{+2}_{-1}$ & $1.14$ & $143^{+2}_{-2}$ &  $217^{+2}_{-2}$  & $1.11$ \\
$151^{+2}_{-2}$ & $209^{+2}_{-2}$ & $-0.87$ & $153^{+1}_{-3}$ & $297^{+3}_{-1}$  & $-0.87$\\
$212^{+4}_{-1}$ & $148^{+1}_{-4}$ & $-0.83$ & $210^{+4}_{-1}$ & $150^{+1}_{-4}$  &  $-0.68$\\
$249^{+5}_{-5}$ & $111^{+5}_{-5}$ & $4.55$ & $246^{+5}_{-5}$ &  $114^{+5}_{-5}$  &  $4.60$\\
$313^{+1}_{-5}$ & $47^{+5}_{-1}$ & $-1.91$ & $310^{+2}_{-5}$ &  $50^{+5}_{-2}$  &  $-1.85$\\
$322^{+2}_{-6}$ & $38^{+6}_{-2}$ & $2.01$ & $320^{+2}_{-7}$ &   $40^{+7}_{-2}$  &  $1.90$\\
$326^{+1}_{-1}$ & $34^{+1}_{-1}$ & $-0.37$ & $324^{+1}_{-1}$ &  $36^{+1}_{-1}$  &  $-0.42$ \\
$328^{+1}_{-1}$ & $32^{+1}_{-1}$ & $0.35$ & $326^{+1}_{-1}$ &   $34^{+1}_{-1}$  &  $0.32$\\
\hline
\end{tabular}
\tnotecaption{The different cases (Fig.~\ref{fig:cases}) are indicated by the indices.
Values for $\alpha_3$ and $\alpha_2$ are deduced from Eq.~\ref{eq:alpha_cases}.}
\end{table}
For the given intervals, we consider all $\Delta\chi^2$ values aside the maxima and minima, whose values are larger than $\pm0.15$.
Thus, this interval is not to be understood as a conventional error margin of $\alpha$, but rather as an interval where the $\chi^2$ is changed significantly.
The results for the two remaining cases, 2 and 3, are simply calculated by using 
Eq.~\ref{eq:alpha_cases}.

\begin{figure}
\centering
\includegraphics[width=0.48\textwidth]{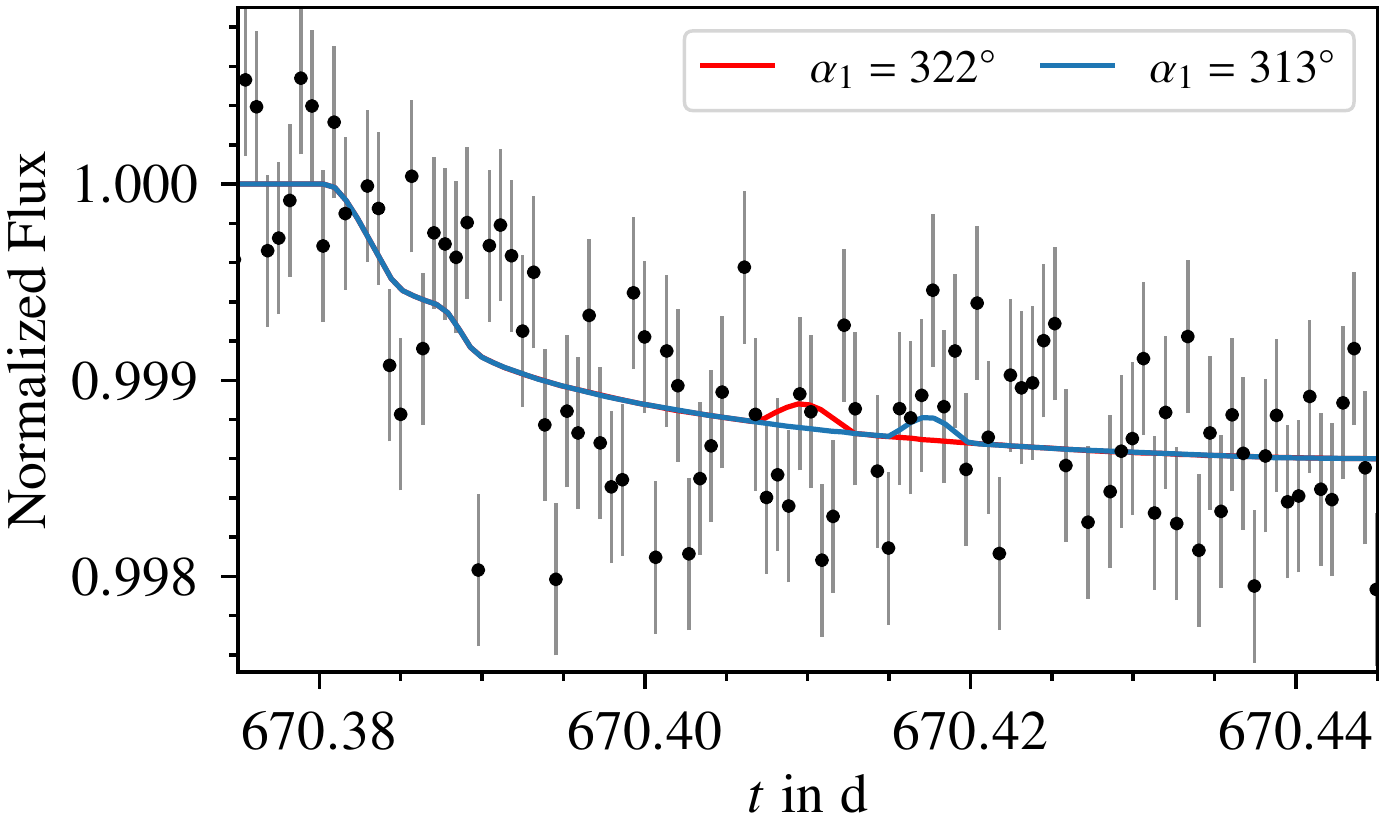}
 \caption{Cutout of our multiple transit number 2. \textit{Dots} show the observational data with $1\sigma$ errors.
 \textit{Solid lines} represent Blender models simulated for different values of $\alpha$ (\textit{labels}).}
  \label{fig:t2_mm} 
\end{figure}

We can see in Table~\ref{tab:angles} that the mentioned shift of $\alpha$ between case 1 and 4 is only $2\degr$ to $3\degr$ and that the differences in $\Delta\chi^2$ between these cases are not very pronounced, especially if the scatter of $\pm0.05$ is considered.
As only one parameter has been varied  in our simulations, a $\Delta\chi^2 = \pm1$ would indicate an influence of this parameter change on the model at the $1\sigma$ level.
As in Fig.~\ref{fig:delta_chi_sq}, we identify four angles that lead to a poorer description of the data and only one angle per case where the model leads to a $\Delta\chi^2 < -1,$ namely $\alpha_1 = 313\degr$ and $\alpha_4=310\degr$.
These angles can be translated to $-47\degr$ and $-50\degr$, respectively.
To further interpret this result, we have to investigate the multiple transit responsible for that dip.
We show a cutout of the multiple transit number 2 in Fig.~\ref{fig:t2_mm} together with Blender models for the angle mentioned above, and $\alpha_1 = 322\degr$ leading to the maximum $\Delta\chi^2$ value in that transit.
Both models clearly differ in the position of the PPO, shifting to earlier times for increasing $\alpha$.
One can see that the PPO of the minimum $\Delta\chi^2$ solution is located where the data points tend to lie above the model, and thus the bump caused by the artificial PPO reduces the $\chi^2$.
In this process, six data points are involved, while the phase where the data points tend to lie above the model lasts longer, roughly from $670.415$\,d to $670.425$\,d.
We believe that this behavior can be attributed to photospheric activity, like a spot crossing event of at least one of the planets, which is more likely than a PPO in that system.
Visual inspection of transits 1 and 5 reveals a similar behavior.

\subsubsection{Favored orbital alignments}
Because of the complete rotational symmetry, we must bear in mind that, for example, case 2 in the retrograde direction is the same as $\alpha_1\pm180\degr$, and is therefore not mentioned in Table~\ref{tab:angles}.
However, we can now investigate fundamental orbital geometries from the different initial cases keeping these symmetries in mind.
Case 1 allows us to simulate the orbital geometry of planets b and c, which is prograde and is, most of the time, in front of the same stellar hemisphere for angles in the interval between $\alpha_1 = [0\degr,\,90\degr]$ and $\alpha_1 = [271\degr,\,359\degr]$. 
In the interval $\alpha_1 = [91\degr,\,270\degr],$ the simulations account for retrograde orbits transiting most of the time in front of different stellar hemispheres.
The situation changes for case 4, where the simulated geometry is prograde and in front of different stellar hemispheres for the intervals $\alpha_4 = [0\degr,\,90\degr]$ and $\alpha_4 = [271\degr,\,359\degr]$, while $\alpha_4 = [91\degr,\,270\degr]$ represent simulations on the same stellar hemisphere in the retrograde direction. 
The latter simulations cover complementary orbital geometries when compared to case 1, opening up the possibility to determine which of these geometries is favored by the data.
To obtain an estimate, we use Eq.~\ref{eq:prob} and integrate the posterior in the intervals mentioned above and summarize the results in Table~\ref{tab:prob_geo}.
\begin{table}
\renewcommand{\arraystretch}{1.25}
\caption[]{Simulated orbital geometries and our probability estimate.}
\label{tab:prob_geo}
\centering
\begin{tabular}{lcc}
\hline\hline
Case and geometry & $p\,/\,\%$ & $\sum(\Delta\chi^2 > 0)$ \\
\hline
1: same hemisphere, prograde & $50.33$ & $35.9$ \\
1: different hemispheres, retrograde & $49.67$ & $39.9$ \\
4: different hemispheres, prograde & $51.09$ & $26.3$ \\
4: same hemisphere, retrograde & $48.91$ & $49.9$\\
\hline
\end{tabular}
\tnotecaption{The probability sums up to $100\,\%$ for each case individually.
The values are calculated from the curves shown in Fig.~\ref{fig:delta_chi_sq} and in the corresponding intervals mentioned in the text.}
\end{table}
In this way, a probability $p<50\,\%$ indicates a geometry that is not favored by the data.
As Eq.~\ref{eq:prob} only depends on $\Delta\chi^2(\alpha)$, simply summing up the $\Delta\chi^2$ values in the corresponding intervals can also serve as an estimate for which geometry is favored.
Larger values of this sum indicate geometries where more angles are rejected by the data or where the values themselves are larger and are thus not favored by the data.
As we can see in Table~\ref{tab:prob_geo}, simulations that lead to a retrograde orbit configuration in cases 1 and 4  show probabilities of slightly smaller than $50\,\%$, while prograde solutions are slightly preferred.
Furthermore, the situation is the same if we neglect negative $\Delta\chi^2$ (Table~\ref{tab:prob_geo}, right column), for example assuming that these values are caused by stellar activity as mentioned above.
Therefore, we argue that according to this analysis the data favor an orbital geometry where both planets eclipse in front of different stellar hemispheres and in the prograde direction.

\section{Discussion}

\subsection{Comparison to previous work}

As already stated above, our transit parameter fits show smaller uncertainties overall when compared to the previous results of \citet{Buchhave2016}.
Our errors might possibly be underestimated because of the fact that we have a lower number of free fit parameters.
\citet{Buchhave2016} present a comprehensive transit analysis including eccentricities, while we insist on using a circular orbit model.
According to the results of  \citet{Buchhave2016}, the orbits have eccentricities close to zero; we therefore argue that a circular model is sufficiently accurate for our analysis. 
In addition,  \citet{Buchhave2016}  fitted LDCs for each planet individually, while we use fixed stellar limb darkening based on our own model prediction mentioned in Sect. \ref{subsec:LD},
a procedure that leads to higher consistency and smaller credibility intervals.
We argue that a fit of the LDCs is only warranted given reasonable  transit-depth-to-noise ratios \citep[Fig.~1.14]{myThesis}, which is definitely not the case for the Kepler-20 planets.
Furthermore, because of the small planet sizes, the fit results obtained by  \citet{Buchhave2016} show a stronger dependence on the choice of the limb darkening, when compared to 
larger planets \citep[Fig.~1.17]{myThesis}.
The strong correlation between the LDCs and the other transit parameters like the radius ratio explains our slightly different fit parameter results.

\subsection{Obliquity sampling}

Our presented obliquity sampling is based on $\chi^2$ values calculated between angle-dependent models and the {\it Kepler} data.
In general, we achieve a $\Delta\chi^2 \neq 0$ whenever the simulation creates artificial PPOs.
This fact is independent of the data.
However, the exact value, and whether $\Delta\chi^2\,$ is greater or less than zero does depend on the data.
Because PPOs appear as bumps in the light curve, the modeled counterparts can only lead to an improvement of the $\chi^2$ in three cases: there is a real PPO, a spot crossing, or a global flux increase.
We cannot distinguish between these cases, and therefore the results for these angles remain inconclusive. However, there is clearly something in the data (Fig.~\ref{fig:t2_mm}).
In contrast, a $\Delta\chi^2\,>\,0$ can be caused by a facula crossing, a global flux decrease, or there is nothing special in the data.
This means that in this case the PPO hypothesis is not supported by the data and the corresponding angles can be ruled out with high confidence.
Although combinations of the cases are possible, like a real PPO and a facula crossing, we think they are unlikely to show a similar magnitude or duration, or to coincide in time.
Another discussion about the existence of a PPO is presented by \citet{Masuda_2014}, who reports a PPO-like event with a much higher signal-to-noise ratio, almost half the misalignment when compared to ours, and finds arguments against the PPO hypothesis.

\begin{figure}
\centering
\includegraphics[width=0.38\textwidth]{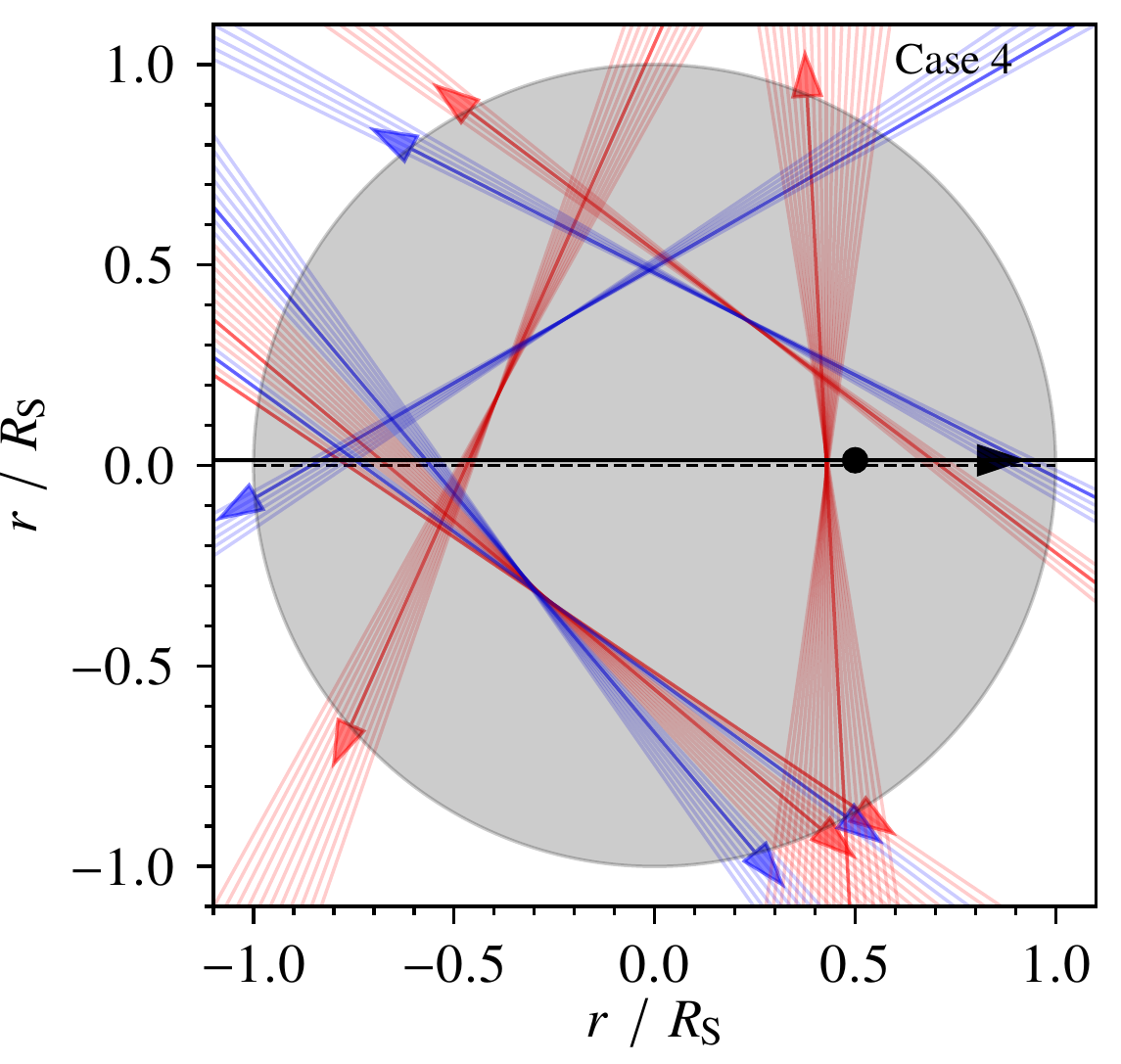}
 \caption{Transit paths of Kepler-20\,b for which simulated PPOs occur.
 Orbital motion indicated by arrows for angles $\alpha_4$ (Table~\ref{tab:angles}).
 Orbits in \textit{red} indicate $\Delta\chi^2>0$, and in \textit{blue} $\Delta\chi^2<0$.
 The stellar equator is marked with a \textit{dashed} line, while planet c is shown as a \textit{black dot}.}
  \label{fig:geo_res} 
\end{figure}

In Fig.~\ref{fig:geo_res}, we visualize the transit paths of  Kepler-20\,b for which we were able to simulate PPOs.
Cases where the PPO hypothesis is not supported by the data ($\Delta\chi^2\,>\,0$) are shown in red, while angles where the results are inconclusive ($\Delta\chi^2\,<\,0$) are shown in blue.
For all other angles we find no artificial PPOs, and therefore we have no information about these geometries.
Nevertheless, this figure nicely illustrates the newly gained information about the orbital configuration when using our method. 

\subsection{Statistical significance of the results}

Unfortunately, our estimate of the most probable orbital geometry (Table~\ref{tab:prob_geo}) is not very distinctive.
The Bayes factor $K,$ as given by \citet{Jeffreys}, calculated for our case 4 leads only to a ``barely worth mentioning'' rating, which means it is not significant. 
However, future investigations of this geometry might show different results.
In this context, it will be particularly interesting to find out whether or not larger planets lead to more significant results.
We carried out basic simulations for larger Kepler-20 planets by increasing $p$ by constant factors in our models and also simply scaling the transit depth of the real data by the same amount.
For this setup, we find a maximum probability of about $56.5\,\%$ for planets enlarged $1.45$ times their real size.
Planets of even greater size do not increase the probability of this geometry.
Artificial PPOs which are then larger than the photometric noise would have been detected in the first place, and information hidden in the noise, such as photometric activity, becomes less important in the determination of the posterior.
This is the case for a planet radius $\ge 3\,R_\mathrm{Earth}$ of the smaller component.
In addition, simulated Kepler-20 data with noise lower than a PPO signature, that is $\le 2\cdot10^{-4}$ if the original planet radii are considered, do show a similar behavior.
Here the mentioned probability decreases for lower noise until the data no longer show the tendency of the geometry.
This is not surprising because, due to the symmetry, the method described in Sect.~\ref{sect:posterior} is only valid if there is at least something in the data, or real significant PPOs. 
In the latter case, this will lead to significant results, while in cases like Kepler-20 this can only serve as a hint.
Despite these limitations, our obliquity sampling is still useful for excluding individual angles $\alpha.$ Also,  for larger planets, the peaks and dips become wider covering a larger fraction of angles $\alpha$, leaving only a small fraction of allowed angles.
This approach also benefits from longer observations covering additional multi-transits at different phase angles.

The obliquity sampling results based on previous transit parameters \citep{Buchhave2016} lead to clearly different $\chi^2$ curves as shown in Fig.~\ref{fig:delta_chi_sq}.
This is because of the sensitivity of the PPO occurrence to the orbital parameters. 
The determined transit parameters must therefore be trustworthy.
In contrast to our results in Table~\ref{tab:prob_geo}, these slightly favor transits in front of the same hemisphere in the prograde direction, also with a ``barely worth mentioning'' rating of the Bayes factor.
This is mainly caused by the prominent dip around $26\degr$, which is deeper for case 1.
However, as stated above, the results show overall higher $\chi^2$ values than those based on our transit fits, which is why we attribute a lower significance to these results than to ours.
For a more comprehensive analysis, one could repeat the obliquity sampling for models whose transit parameters have been varied in the range of their $1\sigma$ uncertainties according to their correlations obtained from the MCMC sampling. 
This would provide a set of $\Delta\chi^2$ curves that can be averaged according to the individual likelihoods of the models.
The computational efforts required mean that this is beyond the scope of this paper.

\subsection{Blender as a tool for transit light-curve synthesis}

When using Blender, the computationally intense ray-tracing approach is the only disadvantage, where the use of GPUs is highly recommended.
The advantages clearly predominate, such as the possibility to include arbitrary stellar limb darkening.
We have not yet used Blender's full potential; as mentioned, it is possible to include stellar active regions, although this would require an adequate spot or faculae model. 
We also tried to achieve full phase curves and secondary eclipses in multi-planet systems.
Thanks to the capability to simulate diffuse Lambertian reflection, both were obtained flawlessly. 
Furthermore, for $b \ne 0,$ we identify reflected planetary day light during primary eclipse and missing light in secondary eclipse, with both having a systematic influence on transit depths. 
In parallel to our analysis, \citet{Martin-Lagarde2020} published a study of this effect in primary transits.
Further investigations of these effects, including planetary textures such as clouds and thermal emission, are of special interest for future photometry, but were beyond the scope of this paper.

\section{Summary and conclusions}

Planetary parameters derived from transit light curves are known to be highly correlated with the stellar limb darkening, as the fit results depend 
on the choice of the limb-darkening description.
We present new fit results of the transit parameters of the planets in the Kepler-20 system using the same limb darkening for all planets, obtained from a stellar model atmosphere specifically generated for Kepler-20.
Using this more reliable and consistent approach, we achieve smaller parameter uncertainties than previously reported for 
the Kepler-20 transiting planets.
Our results for Kepler-20\,b are consistent with previous results, while our results for Kepler-20\,c report a marginally smaller planet ($0.5\,\%$) and a $7\,\%$ larger distance 
to the host star.
In the case of the remaining transiting planets Kepler-20\,d, -e, and -f we find the size  and the semi-major axis of planet d to be consistent 
with the literature, while our approach leads to different values compared to the previous results for the other two planets.

We also carried out an analysis of the stellar activity of Kepler-20 and present a brightness map constructed from the light curve.
We find no long-term active regions, that is, spots, occurring repeatedly after stellar rotations, which could help to determine orbital alignments of the transiting planets.
Consistent with previous results, we find a rotation period of $26.9\,$d and a peak-to-peak activity of $0.1\,\%$ to $0.2\,\%$.

For our transit light-curve synthesis, we use the 3D animation software Blender and demonstrate the usefulness of this software for transit calculations by showing
that the residuals between Blender transits and those generated with the established \texttt{occultquad} routine are of the order of $10^{-7}$. 
We furthermore stress that the use of 3D animation software for light-curve simulations is outstanding for visualization of orbital geometries and has an impact 
on public outreach and other activities.

We present a new method to obtain alignment information of planetary orbits in multi-transiting systems. 
Our approach is useful for systems where conventional methods are not able to achieve valuable results. 
In contrast to other methods, we measure the orbit angle with respect to another orbit in that system rather than in reference to the stellar rotation axis.
We gather this kind of information by simulating multiple transits for all relevant orbit angles leading to artificial PPOs which are then compared to the measured data.
Angles that predict PPOs that are not consistent with the data can be excluded from the geometry.
In this way, we are also able to present an estimate of whether the orbits are aligned in front of the same or different stellar hemispheres, and in which direction 
the planets move in the sense of pro- or retrograde orbital motion.
This estimate becomes significant for real PPOs in the data, with a planet radius $\ge 3\,R_\mathrm{Earth}$ for the smaller component, or for noise 
which is $\le 2\cdot10^{-4}$, if the quality of the photometry or the planet radii are equivalent to our shown example.

As a test scenario and a proof of concept we apply our method to the multiple transits of Kepler-20\,b and c, and find a slight tendency for the planets to orbit in the 
same direction but to transit on different stellar hemispheres ($51.1\,\%$).
Although our results here are not statistically significant, our method allows the exclusion of orbital angles that predict PPOs not consistent with the data.
In this way, this method allows the user to gain insight into angular distributions in multiple systems for which properly observable orbit orientations are still unavailable.
 
\begin{acknowledgements}
This paper includes data collected by NASA's \textit{Kepler} mission. The \textit{Kepler} observations were obtained from the Mikulski Archive for Space Telescopes (MAST). H.\,M.\,M. is supported by the German Research Foundation (DFG) under project number Schm 1032/71-1, and by the company NVIDIA with a GeForce Titan Xp graphics card. Special thanks to T.\,O.\,B.~Schmidt, S.~Czesla, V.\,M.~Passegger, and S.~Dahmke for helpful discussions, and also to P.~Hauschildt for providing a PHOENIX model atmosphere. This work was also supported by COVID-19 financial assistance provided by the DFG.
Additionally, we thank the A\&A Editorial Office.
\end{acknowledgements}

\bibliographystyle{aa}
\bibliography{k20}

\begin{appendix}
 \section{Fit results of remaining Kepler-20 planets}
As outlined in section \ref{sec:methods} we carried out transit parameter fits for all transiting planets in the Kepler-20 system.
Although we do not use the planets d, e, and f in our analysis, here we provide the  parameters that have been found together with their 1$\sigma$ uncertainties, determined as before for planets b and c using MCMC sampling. 
 \begin{table}[h!]
\renewcommand{\arraystretch}{1.25}
\caption[]{Our MCMC transit parameters of Kepler-20\,d, e, and f}
\label{tab:app_paras}
\centering
\begin{tabular}{lccc}
\hline\hline
Parameter & \object{Kepler-20\,d} & \object{Kepler-20\,e}\\
\hline
$R_\mathrm{P}\,/\,R_\mathrm{S}$    & $0.026510^{+0.000263}_{-0.000003}$  & $0.007786^{+0.000063}_{-0.000022}$   \\
$i\,/\,$\degr                      & $89.8876^{+0.0321}_{-0.1396}$       & $89.6241^{+0.0077}_{-0.4867}$        \\
$a\,/\,R_\mathrm{S}$               & $83.196^{+0.026}_{-3.022}$          & $16.341^{+0.015}_{-0.395}$           \\
$T_0\,/\,$d                        & $997.728334^{+0.000275}_{-0.000130}$& $968.934474^{+0.000151}_{-0.000152}$ \\
$P_\mathrm{Orb}\,/\,$d             & $77.611523^{+0.000013}_{-0.000024}$ & $6.098533^{+0.000001}_{-0.000001}$   \\
\hline\hline
Parameter & \object{Kepler-20\,f} \\
\hline
$R_\mathrm{P}\,/\,R_\mathrm{S}$    &   $0.008280^{+0.000092}_{-0.000026}$ &\\
$i\,/\,$\degr                      &   $89.1406^{+0.0005}_{-0.0783}$ &\\
$a\,/\,R_\mathrm{S}$               &   $36.142^{+0.008}_{-0.927}$&\\
$T_0\,/\,$d                        &   $968.205775^{+0.000200}_{-0.000233}$&\\
$P_\mathrm{Orb}\,/\,$d             &   $19.577528^{+0.000005}_{-0.000006}$& \\
\hline
\end{tabular}
\tnotecaption{Time of first transit center is given in BJD$-2.454\cdot10^6$\,d. LDCs fixed to values obtained by a fit to model intensities (Sect.~\ref{subsec:LD}). }
\end{table}

\section{The prototype planet--planet eclipse in Blender}
As probably the most prominent example of PPOs, we mention the prototype of such events, reported and analyzed by \citet{Hirano2012} and \citet{Masuda_2013}.
In a multi-transit of Kepler-89\,d and e, a PPO event occurred near the transit center.
As a test scenario for our transit synthesis, we used the transit parameters from \citet[][their Table~3 and 4]{Masuda_2013}, including a quadratic limb darkening, to reproduce this light curve.
\begin{figure}[h!]
\centering
\includegraphics[width=0.48\textwidth]{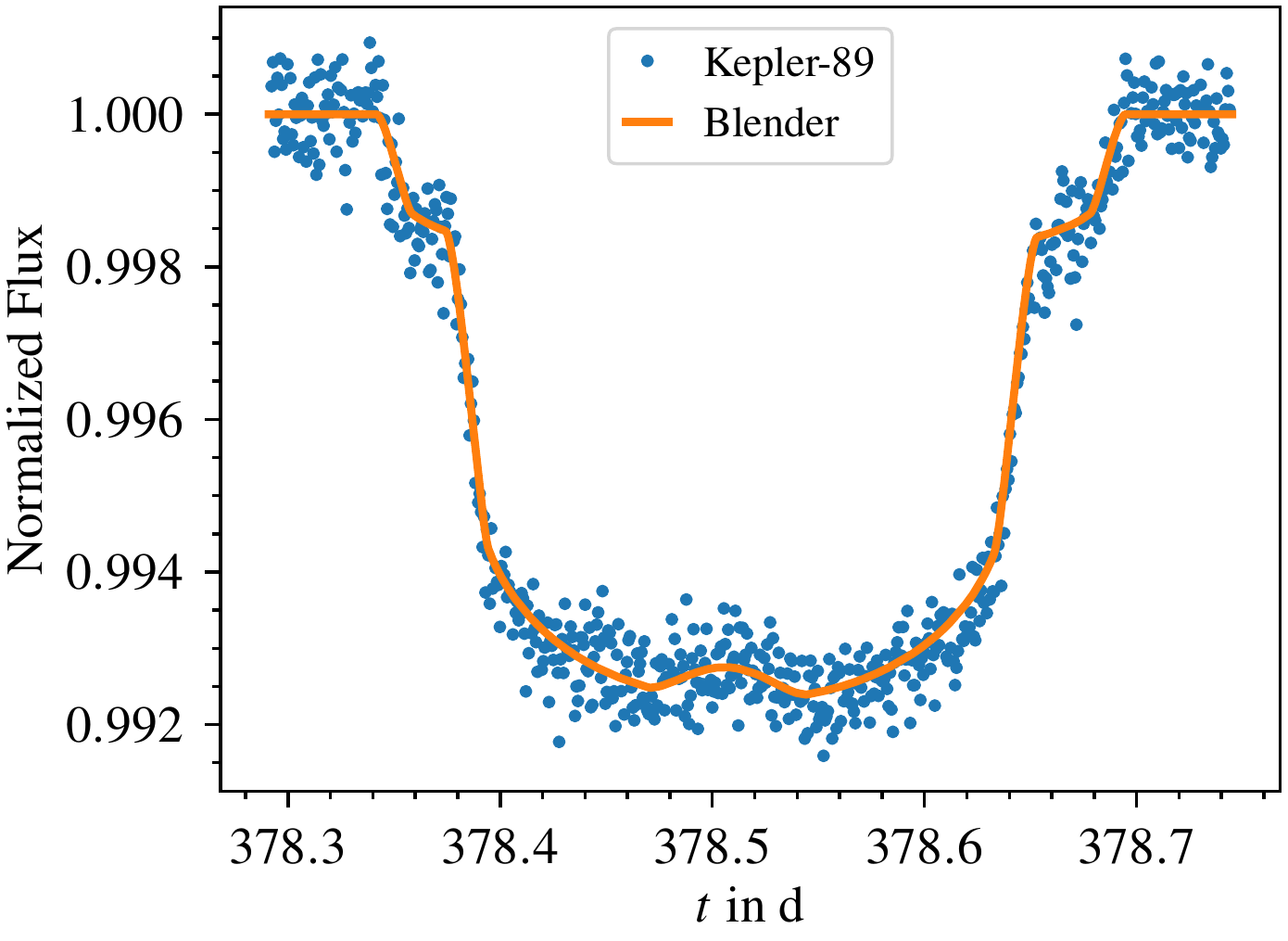}
 \caption{Comparison of the multi-transit light curve Kepler-89\,d\,+\,e (\textit{dots}) with our Blender model with $\alpha = -1.13\degr$ (\textit{solid line}). }
  \label{fig:masuda} 
\end{figure}
The figure shown here presents our resulting Blender light curve, which nicely reproduces the observed transits, including the PPO event. No parameter fits or different limb-darkening models were applied. 
This result underlines the capabilities of the Blender software.
\end{appendix}

\end{document}